# A Unified Theoretical Framework for Mapping Models for the Multi-State Hamiltonian

Jian Liu [1, a)]

*1. Beijing National Laboratory for Molecular Sciences, Institute of Theoretical and Computational Chemistry, College of Chemistry and Molecular Engineering, Peking University, Beijing 100871, China*

a) Electronic mail: jianliupku@pku.edu.cn

We propose a new unified theoretical framework to construct equivalent representations of the multi-state Hamiltonian operator and present several approaches for the mapping onto the Cartesian phase space. After mapping an $F$-dimensional Hamiltonian onto an $F$+1-dimensional space, creation and annihilation operators are defined such that the $F$+1 dimensional space is complete for any combined excitations. Commutation and anti-commutation relations are then naturally derived, which show that the underlying degrees of freedom are neither bosons nor fermions. This sets the scene for developing equivalent expressions of the Hamiltonian operator in quantum mechanics and their classical/semiclassical counterparts. Six mapping models are presented as examples. The framework also offers a novel way to derive such as the well-known Meyer-Miller model.

## I. Introduction

There is considerable effort focused on developing approaches for describing quantum mechanical effects with classical and semiclassical dynamics. The lack of classical analogy with discrete quantum degrees of freedom has presented a challenge in such as non-adiabatic dynamics where multi electronic states are involved. In 1979 Meyer and Miller first proposed a mapping of an $F$-electronic-state Hamiltonian operator onto continuous degrees of freedom[1]. In 1997 Stock and Thoss further presented a more rigorous way to derive the Meyer-Miller Hamiltonian from Schwinger's theory of angular momentum[2]. Since then the Meyer-Miller Hamiltonian has provided a useful theoretical framework to develop (approximate) multi-state quantum dynamics methods.

Although the Meyer-Miller mapping model has successfully been implemented into various examples, it is worth investigating other possible mappings. Cotton and Miller have also pointed out that the Meyer-Miller mapping "is not the most natural one" and proposed a spin mapping model[3]. The purpose of the paper is to present a new theoretical framework to consistently construct equivalent expressions of the multi-state Hamiltonian operator and then yield their mappings onto the Cartesian phase space such that classical dynamics can be implemented. The outline of the paper is as follows. Section II begins by reviewing the application of Schwinger's oscillator model of angular momentum to the multi-state Hamiltonian operator. It then introduces a consistent way to define a pair of creation and annihilation operators in a 2-state space, which naturally leads to their commutation and anti-commutation relations. Section III presents several equivalent representations of the multi-state Hamiltonian operator and produces their classical counterparts. Section IV discusses on the difference between the proposed classical mapping models and the spin mapping model of Ref[3]. Our conclusions are given in Section V.

## II. Multi-state Hamiltonian operator and creation and annihilation operators

### 1. Application of Schwinger's oscillator model of angular momentum

Consider a Hamiltonian operator for $F$ orthonormal states.

$$\hat{H} = \sum_{m,n=1}^{F} H_{nm} \left|n\right\rangle\left\langle m\right| \quad . \tag{1}$$

The Hamiltonian matrix is often a real symmetric one, where $H_{nm} = H_{mn}$. For convenience, the reduced Planck constant is set to $\hbar = 1$ throughout the paper. Stock and Thoss extended Schwinger's oscillator model of angular momentum[4, 5], suggesting that state $|n\rangle$ can be mapped as

$$|n\rangle \mapsto \underbrace{\left|0_1 \cdots 1_n \cdots 0_F\right\rangle}_{F-\text{states}} \tag{2}$$

such that it is viewed as a single excitation from the vacuum state $|\bar{0}\rangle \mapsto \underbrace{\left|0_1 \cdots 0_n \cdots 0_F\right\rangle}_{F-\text{states}}$, i.e.,

$$|n\rangle = \hat{a}_n^{+} |\bar{0}\rangle \quad . \tag{3}$$

Here an excitation represents the occupation of the corresponding state. The vacuum state $|\bar{0}\rangle$ is orthogonal to any occupied state $|n\rangle$. Stock and Thoss followed Schwinger's original oscillator model and represented the $F$ continuous degrees of freedom by the conventional harmonic-oscillator creation and annihilation operators $\left\{\hat{a}_n^{+}, \hat{a}_n\right\}$, commutation relations of which are

$$\left[\hat{a}_m, \hat{a}_n^{+}\right] = \delta_{mn} \quad (\forall m, n) \quad . \tag{4}$$

The multi-state Hamiltonian operator in Eq. (1) is then suggested as[2]

$$\hat{H} = \sum_{m,n=1}^{F} H_{nm} \hat{a}_n^{+} \hat{a}_m \quad . \tag{5}$$

Because a harmonic oscillator contains an infinite number of energy levels, the mapping of Eq. (5) is restricted onto the oscillator subspace with a single excitation, i.e., only levels 0 and 1 are employed. (See more discussion in Appendix A.)

It would be more natural to map each of the $F$ states onto a degree of freedom that has only two states—the vacuum state and the occupied one. A crucial step for constructing such a mapping is to seek a more natural definition of creation and annihilation operators rather than directly apply Schwinger's oscillator model.

**2. Creation and annihilation operators**

As the $F$ states of Eq. (1) are orthonormal, one obtains

$$\begin{aligned} &\langle \bar{0} | \bar{0} \rangle = 1 \\ &\langle n | \bar{0} \rangle = \langle \bar{0} | n \rangle = 0 \\ &\langle m | n \rangle = \delta_{mn} \quad \forall m, n \in \{1, 2, \cdots, F\} \end{aligned} \quad . \tag{6}$$

Since only a single excitation is involved, the creation and annihilation operators can be defined as

$$\begin{aligned} \hat{a}_n^{+} &= |n\rangle \langle \bar{0}| \\ \hat{a}_n &= |\bar{0}\rangle \langle n| \end{aligned} \quad . \tag{7}$$

It is straightforward to verify

$$\hat{a}_n^{+} |n\rangle = 0, \quad \hat{a}_n^{+} |m\rangle = 0 \ \ (n \neq m), \tag{8}$$

$$\hat{a}_n |\bar{0}\rangle = 0, \quad \hat{a}_n |n\rangle = |\bar{0}\rangle, \quad \hat{a}_n |m\rangle = 0 \ \ (n \neq m) \ . \tag{9}$$

Eqs. (6)-(9) suggest a consistent complete space of all excitations. Any combination of two creation operators leads to zero, i.e.,

$$\hat{a}_n^+ \hat{a}_n^+ = \hat{a}_n^+ \hat{a}_m^+ = 0 \qquad . \tag{10}$$

So does any combination of two annihilation operators, i.e.,

$$\hat{a}_n \hat{a}_n = \hat{a}_n \hat{a}_m = 0 \qquad . \tag{11}$$

The combinations of a pair of creation and annihilation operators are

$$\begin{aligned} &\hat{a}_n^+ \hat{a}_n = |n\rangle\langle n| \\ &\hat{a}_n^+ \hat{a}_m = |n\rangle\langle m| \\ &\hat{a}_n \hat{a}_n^+ = \hat{a}_m \hat{a}_m^+ = |\bar{0}\rangle\langle\bar{0}| \\ &\hat{a}_n \hat{a}_m^+ = \hat{a}_m \hat{a}_n^+ = 0 \qquad (n \neq m) \end{aligned} \qquad . \tag{12}$$

Substituting Eq. (12) into Eq. (1) leads to Eq. (5) without any ambiguity.

It is trivial to show the commutation relations

$$\begin{aligned} &\left[\hat{a}_n^+, \hat{a}_n^+\right] = 0 \\ &\left[\hat{a}_n^+, \hat{a}_m^+\right] = 0 \\ &\left[\hat{a}_n, \hat{a}_n\right] = 0 \\ &\left[\hat{a}_n, \hat{a}_m\right] = 0 \\ &\left[\hat{a}_n^+, \hat{a}_n\right] = |n\rangle\langle n| - |\bar{0}\rangle\langle\bar{0}| = \hat{\sigma}_z^{(n)} \\ &\left[\hat{a}_n^+, \hat{a}_m\right] = |n\rangle\langle m| \quad (n \neq m) \end{aligned} \tag{13}$$

and the anti-commutation relations

$$\begin{aligned} &\left[\hat{a}_n^+, \hat{a}_n^+\right]_+ = 0 \\ &\left[\hat{a}_n^+, \hat{a}_m^+\right]_+ = 0 \\ &\left[\hat{a}_n, \hat{a}_n\right]_+ = 0 \\ &\left[\hat{a}_n, \hat{a}_m\right]_+ = 0 \\ &\left[\hat{a}_n^+, \hat{a}_n\right]_+ = |n\rangle\langle n| + |\bar{0}\rangle\langle\bar{0}| = \hat{\mathbf{1}}^{(n)} \\ &\left[\hat{a}_n^+, \hat{a}_m\right]_+ = |n\rangle\langle m| \quad (n \neq m) \end{aligned} \qquad . \tag{14}$$

Here $\hat{\mathbf{1}}^{(n)}$ is the identity operator for state $|n\rangle$. Importantly, Eqs. (13)-(14) suggest that the underlying degrees of freedom in Eq. (5) are *neither bosons nor fermions*.

Eqs. (7)-(14) set the scene for developing mapping models. Below we introduce several approaches for mapping the Hamiltonian operator of Eq. (5) onto the Cartesian phase space such that classical dynamics can be employed.

**III. Equivalent expressions of the multi-state Hamiltonian operator and their mapping models in the phase space**

Define the following operators

$$\begin{aligned} \hat{\sigma}_x^{(n)} &= \hat{a}_n + \hat{a}_n^+ \\ \hat{\sigma}_y^{(n)} &= \frac{\hat{a}_n - \hat{a}_n^+}{i} \end{aligned} \quad . \tag{15}$$

It is easy to show

$$\begin{aligned} i\hat{\sigma}_x^{(n)}\hat{\sigma}_y^{(n)} &= -i\hat{\sigma}_y^{(n)}\hat{\sigma}_x^{(n)} = \hat{\sigma}_z^{(n)} \\ \hat{\sigma}_x^{(n)}\hat{\sigma}_x^{(n)} &= \hat{\sigma}_y^{(n)}\hat{\sigma}_y^{(n)} = \hat{\sigma}_z^{(n)}\hat{\sigma}_z^{(n)} = \hat{\mathbf{1}}^{(n)} \end{aligned} \tag{16}$$

where $\hat{\sigma}_z^{(n)}$ is given in the fifth equation of Eq. (13). The commutation relation is

$$\frac{i}{2}\left[\hat{\sigma}_x^{(n)}, \hat{\sigma}_y^{(n)}\right] = \hat{\sigma}_z^{(n)} \quad , \tag{17}$$

or takes a more general form

$$\frac{i}{2}\left[\hat{\sigma}_a^{(n)}, \hat{\sigma}_b^{(n)}\right] = \varepsilon_{abc}\hat{\sigma}_c^{(n)} \quad . \tag{18}$$

Here the Levi-Civita symbol $\varepsilon_{abc}$ is equal to 1 for cyclic permutations of $xyz$, equal to -1 for anti-cyclic permutations, and equal to zero if index *a* and index *b* are repeated. Similarly, the anti-commutation relations are

$$\frac{1}{2}\left[\hat{\sigma}_a^{(n)}, \hat{\sigma}_b^{(n)}\right]_+ = \delta_{ab}\hat{\mathbf{1}}^{(n)} \quad . \tag{19}$$

Here $\delta_{ab}$ is the Kronecker delta. When $n \neq m$, one further obtains

$$\hat{a}_n^+\hat{a}_m = i\hat{\sigma}_x^{(n)}\hat{\sigma}_y^{(m)} = -i\hat{\sigma}_y^{(n)}\hat{\sigma}_x^{(m)} = \hat{\sigma}_x^{(n)}\hat{\sigma}_x^{(m)} = \hat{\sigma}_y^{(n)}\hat{\sigma}_y^{(m)} \quad (n \neq m) \quad . \tag{20}$$

Note that Eqs. (16)-(19) suggest that $\left\{\hat{\sigma}_x^{(n)}, \hat{\sigma}_y^{(n)}, \hat{\sigma}_z^{(n)}\right\}$ represent Pauli matrices (for a spin $1/2$ particle).

## 1. A mapping model from the analogy with the classical angular momentum

It is trivial to derive from Eqs. (15)-(20)

$$\hat{a}_n^+ \hat{a}_n = \frac{1}{2}\left(\hat{\mathbf{1}}^{(n)} + \frac{i}{2}\left[\hat{\sigma}_x^{(n)}, \hat{\sigma}_y^{(n)}\right]\right) \quad , \tag{21}$$

$$\hat{a}_n^+ \hat{a}_m + \hat{a}_m^+ \hat{a}_n = i\left[\hat{\sigma}_x^{(n)}, \hat{\sigma}_y^{(m)}\right] = -i\left[\hat{\sigma}_y^{(n)}, \hat{\sigma}_x^{(m)}\right] \quad (n \neq m) \quad . \tag{22}$$

The multi-state Hamiltonian operator of Eq. (5) then becomes

$$\hat{H} = \sum_n \frac{1}{2}\left(\hat{\mathbf{1}}^{(n)} + \frac{i}{2}\left[\hat{\sigma}_x^{(n)}, \hat{\sigma}_y^{(n)}\right]\right) H_{nn} + \sum_{n<m} \frac{1}{2}\left(i\left[\hat{\sigma}_x^{(n)}, \hat{\sigma}_y^{(m)}\right] - i\left[\hat{\sigma}_y^{(n)}, \hat{\sigma}_x^{(m)}\right]\right) H_{nm} \, . \tag{23}$$

Eq. (23) is an equivalent expression of Eq. (1) in quantum mechanics.

Because Schwinger's formulation of angular momentum is employed in the mapping—Eqs. (2)-(3), it is natural to seek an analogy with classical angular momentum rather than only stick to Schwinger's original oscillator model. Note that classical angular momentum is defined as

$$\mathbf{L} = \mathbf{x} \times \mathbf{p} \quad . \tag{24}$$

An alternative version of Eq. (24) is

$$x_a p_b - x_b p_a = \varepsilon_{abc} L_c \quad . \tag{25}$$

Here $\{x_a, x_b, x_c\} = \{x, y, z\}$ and $\{p_a, p_b, p_c\} = \{p_x, p_y, p_z\}$. The analogy between Eq. (18) and Eq. (25) suggests a mapping to classical angular momentum

$$\frac{i}{2}\left[\hat{\sigma}_a^{(n)}, \hat{\sigma}_b^{(m)}\right] \mapsto x_a^{(n)} p_b^{(m)} - x_b^{(m)} p_a^{(n)} \quad (n \neq m) \tag{26}$$

and

$$\frac{1}{2}\left(\hat{\mathbf{1}}^{(n)}+\frac{i}{2}\left[\hat{\sigma}_x^{(n)},\hat{\sigma}_y^{(n)}\right]\right) \mapsto x^{(n)}p_y^{(n)}-y^{(n)}p_x^{(n)} \quad . \tag{27}$$

Note that $\frac{1}{2}\left(\hat{\mathbf{1}}^{(n)}+\frac{i}{2}\left[\hat{\sigma}_x^{(n)},\hat{\sigma}_y^{(n)}\right]\right)$ instead of $\frac{i}{2}\left[\hat{\sigma}_x^{(n)},\hat{\sigma}_y^{(n)}\right]$ is employed in the left-hand side of Eq. (27). This is because that spin does not depend on spatial coordinates and has no classical analog. $\frac{1}{2}\left(\hat{\mathbf{1}}^{(n)}+\hat{\sigma}_z^{(n)}\right)$ rather than $\hat{\sigma}_z^{(n)}$ is more natural to be treated as the z-axis component of a quantum spatial angular momentum operator such that its analog leads to a classical angular momentum. (Here Eq. (17) is used.) The eigenvalues of the $2\times 2$ matrix for $\frac{1}{2}\left(\hat{\mathbf{1}}^{(n)}+\hat{\sigma}_z^{(n)}\right)$ are 0 and 1, which indicates that the operator represents the z-axis component of an angular momentum ($L_z$) which takes either 0 or 1. 0 when the state is unoccupied and 1 when it is occupied.

Inserting Eq. (26) and Eq. (27) into Eq. (23) produces the classical Hamiltonian

$$\begin{aligned} H &= \sum_{n=1}^{F} H_{nn}\left(x^{(n)}p_y^{(n)}-y^{(n)}p_x^{(n)}\right) \\ &+\sum_{n<m} H_{nm}\left[\left(x^{(n)}p_y^{(m)}-y^{(n)}p_x^{(m)}\right)+\left(x^{(m)}p_y^{(n)}-y^{(m)}p_x^{(n)}\right)\right] \end{aligned} \quad . \tag{28}$$

The simplified form is then

$$H = \sum_{n,m=1}^{F} H_{nm}\left(x^{(n)}p_y^{(m)}-y^{(n)}p_x^{(m)}\right) \quad . \tag{29}$$

Its Hamilton's equations of motion produce

$$\begin{aligned} \dot{x}^{(n)} &= \frac{\partial H}{\partial p_x^{(n)}} = -H_{nn}y^{(n)}-\sum_{m\neq n}H_{nm}y^{(m)} \\ \dot{y}^{(n)} &= \frac{\partial H}{\partial p_y^{(n)}} = H_{nn}x^{(n)}+\sum_{m\neq n}H_{nm}x^{(m)} \\ \dot{p}_x^{(n)} &= -\frac{\partial H}{\partial x^{(n)}} = -H_{nn}p_y^{(n)}-\sum_{m\neq n}H_{nm}p_y^{(m)} \\ \dot{p}_y^{(n)} &= -\frac{\partial H}{\partial y^{(n)}} = H_{nn}p_x^{(n)}+\sum_{m\neq n}H_{nm}p_x^{(m)} \end{aligned} \quad , \tag{30}$$

which conserves the classical Hamiltonian Eq. (29) by definition. Eqs. (29)-(30) are noted Model I in the paper. It is easy to show that the sum of occupation numbers is a constant of motion for Eq. (30), i.e.,

$$\frac{d}{dt}\sum_{n=1}^{F}\left(x^{(n)}p_y^{(n)}-y^{(n)}p_x^{(n)}\right)=0 \quad . \tag{31}$$

Interestingly, Model I is reminiscent of the semiclassical second-quantized many-electron Hamiltonian (with only 1-electron interactions) proposed by Li and Miller[6]. This kind of analogy suggests that it is possible to obtain a subtle connection between the mapping model for the multi-state Hamiltonian and that for the many-electron Hamiltonian, as will be discussed in our future work. (Note that electrons are fermions.)

## 2. A mapping model from the analogy with the quantum-classical correspondence for two non-commutable operators

Eq. (15) leads to

$$\begin{aligned}\hat{a}_n^{+} &= \frac{1}{2}\left(\hat{\sigma}_x^{(n)}-i\hat{\sigma}_y^{(n)}\right)\\ \hat{a}_n &= \frac{1}{2}\left(\hat{\sigma}_x^{(n)}+i\hat{\sigma}_y^{(n)}\right)\end{aligned} \quad . \tag{32}$$

It is straightforward to obtain

$$\hat{a}_n^{+}\hat{a}_n=\frac{1}{4}\left(\hat{\sigma}_x^{(n)}\hat{\sigma}_x^{(n)}+\hat{\sigma}_y^{(n)}\hat{\sigma}_y^{(n)}\right)+\frac{i}{4}\left[\hat{\sigma}_x^{(n)},\hat{\sigma}_y^{(n)}\right] \tag{33}$$

$$\hat{a}_n^{+}\hat{a}_m=\frac{1}{4}\left(\hat{\sigma}_x^{(n)}\hat{\sigma}_x^{(m)}+\hat{\sigma}_y^{(n)}\hat{\sigma}_y^{(m)}\right)+\frac{i}{4}\left[\hat{\sigma}_x^{(n)},\hat{\sigma}_y^{(m)}\right] \qquad (n\neq m) \tag{34}$$

or

$$\begin{aligned}\hat{a}_n^{+}\hat{a}_m+\hat{a}_m^{+}\hat{a}_n &= \frac{1}{4}\left(\left[\hat{\sigma}_x^{(n)},\hat{\sigma}_x^{(m)}\right]_{+}+\left[\hat{\sigma}_y^{(n)},\hat{\sigma}_y^{(m)}\right]_{+}\right)\\ &\quad +\frac{i}{4}\left(\left[\hat{\sigma}_x^{(n)},\hat{\sigma}_y^{(m)}\right]+\left[\hat{\sigma}_x^{(m)},\hat{\sigma}_y^{(n)}\right]\right) \qquad (n\neq m)\end{aligned} \quad . \tag{35}$$

The multi-state Hamiltonian operator of Eq. (5) then becomes

$$\hat{H} = \sum_n \left( \frac{1}{4}\left( \hat{\sigma}_x^{(n)}\hat{\sigma}_x^{(n)} + \hat{\sigma}_y^{(n)}\hat{\sigma}_y^{(n)} \right) + \frac{i}{4}\left[ \hat{\sigma}_x^{(n)}, \hat{\sigma}_y^{(n)} \right] \right) H_{nn}$$
$$+ \sum_{n<m} \frac{1}{4}\left( \left[ \hat{\sigma}_x^{(n)}, \hat{\sigma}_x^{(m)} \right]_+ + \left[ \hat{\sigma}_y^{(n)}, \hat{\sigma}_y^{(m)} \right]_+ \right) H_{nm} \tag{36}$$
$$+ \sum_{n<m} \frac{i}{4}\left( \left[ \hat{\sigma}_x^{(n)}, \hat{\sigma}_y^{(m)} \right] + \left[ \hat{\sigma}_x^{(m)}, \hat{\sigma}_y^{(n)} \right] \right) H_{nm}$$

Eq. (36) is another equivalent expression of Eq. (1) in quantum mechanics.

Recall the conventional quantum-classical correspondence for two non-commutable operators $\hat{A}$ and $\hat{B}$

$$\begin{aligned} &\frac{1}{2}\left( \hat{A}\hat{B} + \hat{B}\hat{A} \right) \mapsto A(x,p)B(x,p) \\ &[\hat{A}, \hat{B}] \mapsto 0 \end{aligned} \quad . \tag{37}$$

Applying Eq. (37) to Eq. (36) leads to the classical Hamiltonian

$$H = \sum_n \frac{1}{4}\left( \sigma_x^{(n)}\sigma_x^{(n)} + \sigma_y^{(n)}\sigma_y^{(n)} \right) H_{nn} + \sum_{n<m} \frac{1}{2}\left( \sigma_x^{(n)}\sigma_x^{(m)} + \sigma_y^{(n)}\sigma_y^{(m)} \right) H_{nm} \quad . \tag{38}$$

Making a change of variables

$$\begin{aligned} x^{(n)} &= \frac{\sigma_x^{(n)}}{\sqrt{2}} \\ p^{(n)} &= \frac{\sigma_y^{(n)}}{\sqrt{2}} \end{aligned} \tag{39}$$

in Eq. (38) finally yields

$$H = \sum_{n,m=1}^{F} \frac{1}{2}\left( x^{(n)}x^{(m)} + p^{(n)}p^{(m)} \right) H_{nm} \quad . \tag{40}$$

Its Hamilton's equations of motion are then

$$\begin{aligned} \dot{x}^{(n)} &= \frac{\partial H}{\partial p^{(n)}} = H_{nn}p^{(n)} + \sum_{m\neq n} H_{nm}p^{(m)} \\ \dot{p}^{(n)} &= -\frac{\partial H}{\partial x^{(n)}} = -H_{nn}x^{(n)} - \sum_{m\neq n} H_{nm}x^{(m)} \end{aligned} \quad , \tag{41}$$

which preserves the classical Hamiltonian of Eq. (40). It is trivial to verify that the sum of occupation numbers is a constant of motion for Eq. (41), i.e.,

$$\frac{d}{dt}\sum_{n=1}^{F}\frac{\left(x^{(n)}\right)^2+\left(p^{(n)}\right)^2}{2}=0 \quad . \tag{42}$$

Eqs. (40)-(41) are noted Model II in the paper.

If semiclassical/quasiclassical dynamics is employed, the commutation relations in Eq. (36) may not be ignored. Using the parameters $\gamma$ and $\eta$ to describe the effects of $\frac{i}{4}\left[\hat{\sigma}_x^{(n)},\hat{\sigma}_y^{(n)}\right]$ and of $\frac{i}{4}\left(\left[\hat{\sigma}_x^{(n)},\hat{\sigma}_y^{(m)}\right]+\left[\hat{\sigma}_x^{(m)},\hat{\sigma}_y^{(n)}\right]\right)$, respectively, one then obtains the Hamiltonian

$$H_{SC}=\sum_n\frac{1}{2}\left(\left(x^{(n)}\right)^2+\left(p^{(n)}\right)^2-\gamma\right)H_{nn}+\sum_{n<m}\left(x^{(n)}x^{(m)}+p^{(n)}p^{(m)}-\eta\right)H_{nm} \quad . \tag{43}$$

For instance, because $1/2$ is an eigenvalue of the operator $\frac{i}{4}\left[\hat{\sigma}_x^{(n)},\hat{\sigma}_y^{(n)}\right]=\frac{\hat{\sigma}_z^{(n)}}{2}$, the value of the parameter $\gamma$ can be chosen as $1/2$. More generally, an optimum value for $\gamma$ can be selected in the regime between the two eigenvalues, i.e., $\left[-1/2,1/2\right]$. When the commutation relations are not taken into account (i.e., $\gamma=0$ and $\eta=0$), Eq. (43) approaches the classical mapping Hamiltonian Eq. (40).

The classical Hamiltonian in Eq. (40) in Model II or the semiclassical one in Eq. (43) is closely related to the well-known Meyer-Miller Hamiltonian

$$H_{MM}=\sum_n\frac{1}{2}\left(\left(x^{(n)}\right)^2+\left(p^{(n)}\right)^2-\gamma\right)H_{nn}+\sum_{n<m}\left(x^{(n)}x^{(m)}+p^{(n)}p^{(m)}\right)H_{nm} \tag{44}$$

where the parameter $\gamma$ is set to $1/2$ in Meyer and Miller's original version[1, 2] or chosen to be $\left(\sqrt{3}-1\right)/2$ or other optimal values in its applications[7-11]. The derivation procedure of Eq. (40) or Eq. (43) is very different from Meyer and Miller's or from Stock and Thoss' seminal work though[1, 2, 12]. As long as creation and annihilation operators are defined in Eq. (7), Eq. (13) demonstrates that conventional harmonic-oscillator commutation relations Eq. (4) do *not* hold.

## 3. Three mapping models from the analogy with the classical vector

Note that the Pauli matrix $\hat{\sigma}_a$ in physics represents the observables corresponding to spin along the $a$-th coordinate axis. When only $\hat{\sigma}_x$ and $\hat{\sigma}_y$ appear in the Hamiltonian operator, it is natural to adopt the mapping of the two Pauli vectors onto two 2-dimensional vectors

$$\hat{\sigma}_a^{(n)}\hat{\sigma}_b^{(m)} \mapsto \left(x_a^{(n)}+ip_a^{(n)}\right)\left(x_b^{(m)}-ip_b^{(m)}\right) \quad . \tag{45}$$

The mapping Eq. (45) leads to

$$\begin{aligned} &\hat{\sigma}_a^{(n)}\hat{\sigma}_a^{(n)} = \frac{\left[\hat{\sigma}_a^{(n)},\hat{\sigma}_a^{(n)}\right]_+}{2} \mapsto \left(x_a^{(n)}\right)^2+\left(p_a^{(n)}\right)^2 \\ &\frac{\left[\hat{\sigma}_a^{(n)},\hat{\sigma}_a^{(m)}\right]_+}{2} \mapsto x_a^{(n)}x_a^{(m)}+p_a^{(n)}p_a^{(m)} \quad (n\neq m) \end{aligned} \tag{46}$$

and

$$\begin{aligned} &\frac{i}{2}\left[\hat{\sigma}_x^{(n)},\hat{\sigma}_y^{(n)}\right] \mapsto x^{(n)}p_y^{(n)}-y^{(n)}p_x^{(n)} \\ &\frac{i}{2}\left[\hat{\sigma}_x^{(n)},\hat{\sigma}_y^{(m)}\right] \mapsto x^{(n)}p_y^{(m)}-y^{(m)}p_x^{(n)} \quad (n\neq m) \end{aligned} \quad . \tag{47}$$

### 3-1. Model III

Substituting Eq. (19) into Eq (23) produces the 3rd equivalent representation of the multi-state Hamiltonian operator in quantum mechanics

$$\begin{aligned} \hat{H} = &\sum_n \frac{1}{4}\left(\hat{\sigma}_x^{(n)}\hat{\sigma}_x^{(n)}+\hat{\sigma}_y^{(n)}\hat{\sigma}_y^{(n)}+i\left[\hat{\sigma}_x^{(n)},\hat{\sigma}_y^{(n)}\right]\right)H_{nn} \\ &+\sum_{n<m}\frac{1}{2}\left(i\left[\hat{\sigma}_x^{(n)},\hat{\sigma}_y^{(m)}\right]-i\left[\hat{\sigma}_y^{(n)},\hat{\sigma}_x^{(m)}\right]\right)H_{nm} \end{aligned} \quad . \tag{48}$$

Applying the mapping Eqs. (46) and (47) to Eq. (48) yields the classical Hamiltonian

$$H = \sum_{n=1}^{F}\frac{\left(x^{(n)}+p_y^{(n)}\right)^2+\left(y^{(n)}-p_x^{(n)}\right)^2}{4}H_{nn}+\sum_{n\neq m}\left(x^{(n)}p_y^{(m)}-y^{(m)}p_x^{(n)}\right)H_{nm} \quad . \tag{49}$$

Hamilton's equations of motion then read

$$
\begin{aligned}
\dot{x}^{(n)} &= \frac{\partial H}{\partial p_x^{(n)}} = H_{nn} \frac{\left(p_x^{(n)} - y^{(n)}\right)}{2} - \sum_{m \neq n} H_{mn} y^{(m)} \\
\dot{y}^{(n)} &= \frac{\partial H}{\partial p_y^{(n)}} = H_{nn} \frac{\left(x^{(n)} + p_y^{(n)}\right)}{2} + \sum_{m \neq n} H_{mn} x^{(m)} \\
\dot{p}_x^{(n)} &= -\frac{\partial H}{\partial x^{(n)}} = -H_{nn} \frac{\left(x^{(n)} + p_y^{(n)}\right)}{2} - \sum_{m \neq n} H_{mn} p_y^{(m)} \\
\dot{p}_y^{(n)} &= -\frac{\partial H}{\partial y^{(n)}} = H_{nn} \frac{\left(p_x^{(n)} - y^{(n)}\right)}{2} + \sum_{m \neq n} H_{mn} p_x^{(m)}
\end{aligned} \quad , \qquad (50)
$$

which preserves the classical Hamiltonian of Eq. (49). Eqs. (49)-(50) are noted Model III in the paper. It is straightforward to verify conservation of the sum of occupation numbers for Eq. (50), i.e.,

$$
\frac{d}{dt} \sum_{n=1}^{F} \frac{1}{4} \left( \left( x^{(n)} + p_y^{(n)} \right)^2 + \left( y^{(n)} - p_x^{(n)} \right)^2 \right) = 0 \quad . \qquad (51)
$$

### 3-2. Model IV

Similarly, Model IV is constructed from the 4-th equivalent representation of the multi-state Hamiltonian operator

$$
\begin{aligned}
\hat{H} = &\sum_n \frac{1}{4} \left( \hat{\sigma}_x^{(n)} \hat{\sigma}_x^{(n)} + \hat{\sigma}_y^{(n)} \hat{\sigma}_y^{(n)} + i \left[ \hat{\sigma}_x^{(n)}, \hat{\sigma}_y^{(n)} \right] \right) H_{nn} \\
&+ \sum_{n<m} \frac{1}{2} \left( \left[ \hat{\sigma}_x^{(n)}, \hat{\sigma}_x^{(m)} \right]_+ + \left[ \hat{\sigma}_y^{(n)}, \hat{\sigma}_y^{(m)} \right]_+ \right) H_{nm}
\end{aligned} \quad . \qquad (52)
$$

Note that Eq. (20) leads to

$$
\begin{aligned}
\hat{a}_n^+ \hat{a}_m + \hat{a}_m^+ \hat{a}_n &= \frac{1}{2} \left( i \left[ \hat{\sigma}_x^{(n)}, \hat{\sigma}_y^{(m)} \right] - i \left[ \hat{\sigma}_y^{(n)}, \hat{\sigma}_x^{(m)} \right] \right) \\
&= \frac{1}{2} \left( \left[ \hat{\sigma}_x^{(n)}, \hat{\sigma}_x^{(m)} \right]_+ + \left[ \hat{\sigma}_y^{(n)}, \hat{\sigma}_y^{(m)} \right]_+ \right) \qquad (n \neq m)
\end{aligned} \quad . \qquad (53)
$$

Eq. (52) is then derived from Eq. (53) and Eq. (48). Substituting Eqs. (46) and (47) into Eq. (52) yield the classical Hamiltonian of Model IV

$$
\begin{aligned}
H &= \sum_{n=1}^{F} \frac{\left(x^{(n)} + p_y^{(n)}\right)^2 + \left(y^{(n)} - p_x^{(n)}\right)^2}{4} H_{nn} \\
&+ \sum_{n<m} \left( \left( x^{(n)} x^{(m)} + p_x^{(n)} p_x^{(m)} \right) + \left( y^{(n)} y^{(m)} + p_y^{(n)} p_y^{(m)} \right) \right) H_{nm}
\end{aligned} \quad . \qquad (54)
$$

Hamilton's equations of motion become

$$
\begin{aligned}
\dot{x}^{(n)} &= \frac{\partial H}{\partial p_x^{(n)}} = H_{nn} \frac{\left(p_x^{(n)} - y^{(n)}\right)}{2} + \sum_{m \neq n} H_{nm} p_x^{(m)} \\
\dot{y}^{(n)} &= \frac{\partial H}{\partial p_y^{(n)}} = H_{nn} \frac{\left(x^{(n)} + p_y^{(n)}\right)}{2} + \sum_{m \neq n} H_{nm} p_y^{(m)} \\
\dot{p}_x^{(n)} &= -\frac{\partial H}{\partial x^{(n)}} = -H_{nn} \frac{\left(x^{(n)} + p_y^{(n)}\right)}{2} - \sum_{m \neq n} H_{nm} x^{(m)} \\
\dot{p}_y^{(n)} &= -\frac{\partial H}{\partial y^{(n)}} = H_{nn} \frac{\left(p_x^{(n)} - y^{(n)}\right)}{2} - \sum_{m \neq n} H_{nm} y^{(m)}
\end{aligned} \quad , \qquad (55)
$$

which conserves the classical Hamiltonian of Eq. (54). Conservation of the sum of occupation numbers for Eq. (55) can easily be verified, which shares the same expression as Eq. (51).

### 3-3. Model V

When the equivalent representation of the multi-state Hamiltonian operator in Eq. (36) is used, employing the mapping Eqs. (46) and (47) for Eq. (36) then leads to the classical Hamiltonian of Model V,

$$
\begin{aligned}
H &= \sum_{n=1}^{F} \frac{\left(x^{(n)} + p_y^{(n)}\right)^2 + \left(y^{(n)} - p_x^{(n)}\right)^2}{4} H_{nn} \\
&+ \sum_{n<m} \frac{\left(x^{(n)} + p_y^{(n)}\right)\left(x^{(m)} + p_y^{(m)}\right) + \left(y^{(n)} - p_x^{(n)}\right)\left(y^{(m)} - p_x^{(m)}\right)}{2} H_{nm}
\end{aligned} \quad . \qquad (56)
$$

Its Hamilton's equations of motion are

$$\dot{x}^{(n)} = \frac{\partial H}{\partial p_x^{(n)}} = H_{nn} \frac{\left(p_x^{(n)} - y^{(n)}\right)}{2} + \sum_{m \neq n} H_{nm} \frac{\left(p_x^{(m)} - y^{(m)}\right)}{2}$$

$$\dot{y}^{(n)} = \frac{\partial H}{\partial p_y^{(n)}} = H_{nn} \frac{\left(x^{(n)} + p_y^{(n)}\right)}{2} + \sum_{m \neq n} H_{nm} \frac{\left(x^{(m)} + p_y^{(m)}\right)}{2}$$

$$\dot{p}_x^{(n)} = -\frac{\partial H}{\partial x^{(n)}} = -H_{nn} \frac{\left(x^{(n)} + p_y^{(n)}\right)}{2} - \sum_{m \neq n} H_{nm} \frac{\left(x^{(m)} + p_y^{(m)}\right)}{2} \quad , \quad (57)$$

$$\dot{p}_y^{(n)} = -\frac{\partial H}{\partial y^{(n)}} = H_{nn} \frac{\left(p_x^{(n)} - y^{(n)}\right)}{2} + \sum_{m \neq n} H_{nm} \frac{\left(p_x^{(m)} - y^{(m)}\right)}{2}$$

which conserves the classical Hamiltonian of Eq. (56). Conservation of the sum of occupation numbers for Eq. (57) shares the same expression as Eq. (51).

## 4. Other equivalent representations of the multi-state Hamiltonian operator

The theoretical framework (presented in Section II and in Section III 1-3) yield four equivalent representations (Eqs. (23), (36), (48), and (52)) of the multi-state Hamiltonian operator (Eq. (1) or Eq. (5)) in quantum mechanics. All these equivalent representations are expressed in terms of $\left\{\hat{\sigma}_x^{(n)}, \hat{\sigma}_y^{(n)}\right\}$ in the theoretical framework.

More equivalent representations can be proposed as well. For instance, substituting Eqs. (21), and (53) into Eq. (5) produces the fifth equivalent representation of the multi-state Hamiltonian operator in quantum mechanics

$$\hat{H} = \sum_n \frac{1}{2}\left(\hat{\mathbf{1}}^{(n)} + \frac{i}{2}\left[\hat{\sigma}_x^{(n)}, \hat{\sigma}_y^{(n)}\right]\right) H_{nn} + \sum_{n<m} \frac{1}{2}\left(\left[\hat{\sigma}_x^{(n)}, \hat{\sigma}_x^{(m)}\right]_+ + \left[\hat{\sigma}_y^{(n)}, \hat{\sigma}_y^{(m)}\right]_+\right) H_{nm} \quad . \quad (58)$$

Applying the similar strategies (introduced in the previous part of the section) to Eq. (58) leads to the classical Hamiltonian of Model VI in the Cartesian phase space

$$H = \sum_{n=1}^{F} H_{nn}\left(x^{(n)} p_y^{(n)} - y^{(n)} p_x^{(n)}\right) + \sum_{n<m}\left(\left(x^{(n)} x^{(m)} + p_x^{(n)} p_x^{(m)}\right) + \left(y^{(n)} y^{(m)} + p_y^{(n)} p_y^{(m)}\right)\right) H_{nm} \tag{59}$$

Its equations of motion become

$$\begin{aligned}
\dot{x}^{(n)} &= \frac{\partial H}{\partial p_x^{(n)}} = -H_{nn} y^{(n)} + \sum_{m \neq n} H_{nm} p_x^{(m)} \\
\dot{y}^{(n)} &= \frac{\partial H}{\partial p_y^{(n)}} = H_{nn} x^{(n)} + \sum_{m \neq n} H_{nm} p_y^{(m)} \\
\dot{p}_x^{(n)} &= -\frac{\partial H}{\partial x^{(n)}} = -H_{nn} p_y^{(n)} - \sum_{m \neq n} H_{nm} x^{(m)} \\
\dot{p}_y^{(n)} &= -\frac{\partial H}{\partial y^{(n)}} = H_{nn} p_x^{(n)} - \sum_{m \neq n} H_{nm} y^{(m)}
\end{aligned} \quad . \tag{60}$$

Conservation of the sum of occupation numbers for Eq. (55) can easily be verified, which shares the same expression as Eq. (31).

Similarly, more quantum-classical analogies or more classical mapping models in the Cartesian phase space can be proposed in the theoretical framework.

## IV. Spin mapping model of Cotton and Miller

Note that Eq. (17) leads to an equivalent expression of Eq. (58)

$$\hat{H} = \sum_{n} \frac{1}{2}\left(\hat{\mathbf{1}}^{(n)} + \hat{\sigma}_z^{(n)}\right) H_{nn} + \sum_{n<m} \frac{1}{2}\left(\left[\hat{\sigma}_x^{(n)}, \hat{\sigma}_x^{(m)}\right]_+ + \left[\hat{\sigma}_y^{(n)}, \hat{\sigma}_y^{(m)}\right]_+\right) H_{nm} \quad . \tag{61}$$

Employing the transformation

$$\hat{S}_a^{(n)} = \frac{\hat{\sigma}_a^{(n)}}{2} \qquad (a = x, y, \text{or } z) \quad , \tag{62}$$

one obtains an equivalent expression of Eq. (61)

$$\hat{H} = \sum_{n}\left(\frac{1}{2}\hat{\mathbf{1}}^{(n)} + \hat{S}_z^{(n)}\right) H_{nn} + \sum_{n<m}\left(\left[\hat{S}_x^{(n)}, \hat{S}_x^{(m)}\right]_+ + \left[\hat{S}_y^{(n)}, \hat{S}_y^{(m)}\right]_+\right) H_{nm} \quad . \tag{63}$$

If the approximation $\left[\hat{S}_a^{(n)},\hat{S}_a^{(m)}\right]_+ \approx 2\hat{S}_a^{(n)}\hat{S}_a^{(m)}$ is employed, Eq. (63) then becomes

$$\hat{H}=\sum_n\left(\frac{1}{2}\hat{\mathbf{1}}^{(n)}+\hat{S}_z^{(n)}\right)H_{nn}+2\sum_{n<m}\left(\hat{S}_x^{(n)}\hat{S}_x^{(m)}+\hat{S}_y^{(n)}\hat{S}_y^{(m)}\right)H_{nm} \quad , \tag{64}$$

which is the quantum Hamiltonian operator used for the spin mapping model in Ref. [3]. Cotton and Miller replaced the spin operator in Eq. (64) with the classical angular momentum vector

$$\mathbf{S}^{(i)}\equiv\begin{bmatrix}S_x^{(i)}\\S_y^{(i)}\\S_z^{(i)}\end{bmatrix}=\begin{bmatrix}\sqrt{S^2-\left(m^{(i)}\right)^2}\cos\left(q^{(i)}\right)\\\sqrt{S^2-\left(m^{(i)}\right)^2}\sin\left(q^{(i)}\right)\\m^{(i)}\end{bmatrix} \tag{65}$$

and then obtained the spin mapping Hamiltonian[3]

$$H=\sum_i\left(\frac{1}{2}+S_z^{(i)}\right)H_{ii}+2\sum_{i<j}\left(S_x^{(i)}S_x^{(j)}+S_y^{(i)}S_y^{(j)}\right)H_{ij} \quad . \tag{66}$$

Here $\left\{m^{(i)},q^{(i)}\right\}$ are the action-angle variables for the *n*-th degree of freedom, where the parameter $S^2$ is suggested to be the quantum value $S^2=3/4$. The equations of motion read

$$\begin{aligned}\dot{q}^{(i)}&=\frac{\partial H}{\partial m^{(i)}}=H_{ii}-\frac{2m^{(i)}}{\sqrt{S^2-\left(m^{(i)}\right)^2}}\sum_{j\neq i}H_{ij}\sqrt{S^2-\left(m^{(j)}\right)^2}\cos\left(q^{(i)}-q^{(j)}\right)\\\dot{m}^{(i)}&=-\frac{\partial H}{\partial q^{(i)}}=2\sqrt{S^2-\left(m^{(i)}\right)^2}\sum_{j\neq i}H_{ij}\sqrt{S^2-\left(m^{(j)}\right)^2}\sin\left(q^{(i)}-q^{(j)}\right)\end{aligned} \quad , \tag{67}$$

a more compact form[3] of which is

$$\frac{d\mathbf{S}^{(i)}}{dt}=\frac{\partial H}{\partial\mathbf{S}^{(i)}}\times\mathbf{S}^{(i)} \quad . \tag{68}$$

The spin mapping model is then a nonlinear system, which is very different from the quadratic Hamiltonian models (Models I-VI) that are obtained in the theoretical framework. Note that the spin mapping model does not employ an equivalent representation expressed in terms of $\left\{\hat{\sigma}_x^{(n)}, \hat{\sigma}_y^{(n)}\right\}$ for the multi-state Hamiltonian operator (Eq. (1) or Eq. (5)). This is also different from Models I-VI constructed in the theoretical framework.

The spin mapping model [Eqs. (66)-(65)] can *not* be verified to be an exact mapping model of the time-dependent Schrödinger equation. It works well when the coupling terms $\{H_{nm}\}$ $(n \neq m)$ are weak, but fails in the strong coupling region. (See Appendix C.) It is not a surprise. We point out in Section III-1 that spin does not depend on spatial coordinates and has no classical analog. $\hat{S}_z^{(n)}$ is not natural to be treated as the z-axis component of a quantum spatial angular momentum operator. That is, Eq. (65) is not a good analogy. Cotton and Miller have already demonstrated that the spin mapping model [Eqs. (65)-(66)] is less accurate than the Meyer-Miller mapping model in Ref. [3]. This is mostly because that the spin mapping model is not exact even when the nuclear motions are frozen.

**V. Conclusion remarks**

In this paper, we present a new unified theoretical framework to construct equivalent representations of the multi-state Hamiltonian operator and propose several approaches for the mapping onto the Cartesian phase space. Below we list the three key elements,

1) Extend Schwinger's formulation to map the $F$-dimensional Hamiltonian operator onto a $F+1$ dimensional space. (I. e., Eqs. (2)-(3), as first introduced by Stock and Thoss[2].)
2) Define creation and annihilation operators as in Eq. (7) such that the $F+1$ dimensional space is complete for all (combined) excitations. Commutation and anti-commutation relations are then naturally constructed.

3) Derive equivalent representations of the Hamiltonian operator (in terms of $\left\{\hat{\sigma}_x^{(n)}, \hat{\sigma}_y^{(n)}\right\}$) and propose the criteria for mapping them onto the Cartesian phase space such that classical dynamics can be employed.

Three quantum-classical analogies (or criteria) are proposed. Six classical Hamiltonian models (namely Eq. (29), Eq. (40), Eq. (49), Eq. (54), Eq. (56), and Eq. (59)) are then developed as examples. (Similarly, semiclassical/quasiclassical models can also be developed although they are not explicitly shown in the present paper.) Each of Models I-VI involves a Hamiltonian that has only quadratic terms in the Cartesian phase space. It suggests that the six different classical mapping models can lead to exact quantum results if initial conditions are carefully constructed (as discussed in Appendices B and C). Interestingly, Section III-2 presents a novel derivation for the conventional Meyer-Miller model[1, 2, 12].

Although the quantum Hamiltonian operator used in Ref. [3] [i.e., Eq. (64)] is closely related to an equivalent expression of the multi-state Hamiltonian operator [i.e., Eq. (58)] in the theoretical framework, the spin mapping model of Cotton and Miller[3] [Eqs. (65)-(66)] for Eq. (64) does not have the 3rd key element listed above. It can not be verified to be exact. (See Appendix C.)

Finally, we note that the unified framework offers a way to develop more equivalent representations of the multi-state Hamiltonian operator and their classical/semiclassical mapping models that are able to produce exact results. It will be interesting in future work to seek an optimal and economy classical/semiclassical mapping model for studying real complex systems. When the state in Eq. (1) is the electronic state, $H_{nm} = H_{nm}(\mathbf{R})$ are often general functions of the nuclear coordinates $\mathbf{R}$, so that adding the nuclear kinetic energy operator $\frac{1}{2}\hat{\mathbf{p}}^T\mathbf{M}^{-1}\hat{\mathbf{p}}$ to Eq. (1) leads to the whole nuclear-electronic Hamiltonian operator. It will also be interesting to include the nuclear degrees of freedom in the classical mapping models for the multi-state system. Further investigation along these directions is certainly warranted.

**Acknowledgement**

This work was supported by the National Natural Science Foundation of China (NSFC) Grants No. 21373018 and No. 21573007, by the Recruitment Program of Global Experts, by Specialized Research Fund for the Doctoral Program of Higher Education No. 20130001110009, by the Ministry of Science and Technology of China (MOST) Grant No. 2016YFC0202803, and by Special Program for Applied Research on Super Computation of the NSFC-Guangdong Joint Fund (the second phase). We acknowledge the Beijing and Tianjin supercomputer centers for providing computational resources.

## Appendix A: Creation and annihilation operators in terms of eigenstates for a harmonic oscillator

Consider the Hamiltonian for a unit mass and unit frequency harmonic oscillator

$$\hat{H} = \frac{1}{2}\left(\hat{x}^2 + \hat{p}^2\right) \quad . \tag{69}$$

Its eigenstates $\{ |j\rangle \}$ satisfy

$$\hat{H}|j\rangle = \left( j + \frac{1}{2} \right) |j\rangle , \quad j = 0,1,2,\cdots \quad . \tag{70}$$

The system consists of an infinite number of eigenstates. It is heuristic to think how the creation and annihilation operators can be represented in terms of eigenstates.

Consider the lowest $s+1$ eigenstates which span an $(s+1)$–dimensional state space. The expectation value of any physical property of interest is first expressed in the $(s+1)$–dimensional state space, then we take the limit $s \to \infty$ to obtain the correct result. The annihilation operator is given by

$$\hat{a} = |0\rangle\langle 1| + \sqrt{2}|1\rangle\langle 2| + \cdots + \sqrt{s}|s-1\rangle\langle s| \tag{71}$$

and its Hermitian conjugate produces the creation operator

$$\hat{a}^{+} = |1\rangle\langle 0| + \sqrt{2}|2\rangle\langle 1| + \cdots + \sqrt{s}|s\rangle\langle s-1| \quad . \tag{72}$$

The number operator

$$\hat{N} = \hat{a}^{+}\hat{a} \tag{73}$$

becomes

$$\hat{N} = \sum_{j=0}^{s} j | j\rangle\langle j| \quad . \qquad (74)$$

Eqs. (71)-(72) lead to the commutation relation

$$\left[\hat{a}, \hat{a}^{+}\right] = 1 - (s+1)|s\rangle\langle s| \quad . \qquad (75)$$

As $s$ tends to infinity, the second term of the right-hand side of Eq. (75) has no effect when the commutator $\left[\hat{a}, \hat{a}^{+}\right]$ operates on any *physical* state[13]. For example, the expectation of energy for a general physical state $|\Phi\rangle$ is

$$\langle\Phi|\hat{H}|\Phi\rangle = \lim_{s\to\infty}\left[\sum_{j=1}^{s}\left|\langle\Phi| j\rangle\right|^{2}\left(j + \frac{1}{2}\right)\right] \quad . \qquad (76)$$

Note that the expectation value of energy for a physical state is always finite. This ensures that either $|\langle\Phi|s\rangle|^{2} s$ or $|\langle\Phi|s\rangle|^{2}$ must approach zero when $s$ tends to infinity. The expectation value of $\left[\hat{a}, \hat{a}^{+}\right]$ is

$$\langle\Phi|\left[\hat{a}, \hat{a}^{+}\right]|\Phi\rangle = 1 - (s+1)\left|\langle s|\Phi\rangle\right|^{2} \quad . \qquad (77)$$

The second term of the right-hand side of Eq. (77) must vanish as $s$ tends to infinity. That is, the physical-state commutator

$$\left[\hat{a}, \hat{a}^{+}\right]_{p} = 1 \qquad (78)$$

is sufficient when acting on any physical state[13].

When $s$ is finite, the conventional commutation relation (Eq. (78)) do *not* hold. That is to say, if only finite eigenstates are employed in Eqs. (71)-(72), the operators

$$\begin{aligned} \hat{x} &= \frac{\hat{a}+\hat{a}^{+}}{\sqrt{2}} \\ \hat{p} &= \frac{\hat{a}-\hat{a}^{+}}{\sqrt{2}i} \end{aligned} \tag{79}$$

do *not* lead to the well-known commutation relation

$$\left[\hat{x}, \hat{p}\right] = i \quad . \tag{80}$$

**Appendix B: Models I-VI are exact mapping models of the time-dependent Schrödinger equation**

The amplitudes $\{c_n(t)\}$ for being in the different states at time *t* are determined by the standard time-dependent Schrödinger equation (TDSE),

$$i\dot{c}_n(t) = \sum_{m=1}^{F} H_{nm} c_m(t) \quad . \tag{81}$$

Make a change of variables

$$c_n(t) = x^{(n)}(t) + ip^{(n)}(t) \tag{82}$$

where both $x^{(n)}(t)$ and $p^{(n)}(t)$ are real. Substituting Eq. (82) into Eq. (81), one obtains the equations of motion

$$\begin{aligned} \dot{x}^{(n)}(t) &= \sum_{m=1}^{F} H_{nm} p^{(m)}(t) \\ \dot{p}^{(n)}(t) &= -\sum_{m=1}^{F} H_{nm} x^{(m)}(t) \end{aligned} \quad . \tag{83}$$

Eq. (83) is identical to Eq. (41), which is derived from the classical Hamiltonian Eq. (40) in Model II. Model II (Eqs. (40)-(41)) is then an exact mapping model in quantum mechanics, irrespective of that it is derived as a classical mapping model of Eq. (36). We note that the action-angle version of Eq. (83) was already presented in Meyer and Miller's original work[1].

We then consider Model I. Note that Model I has four variables while Model II has only two. Make the change of variables

$$x = q,\ \ y = -p_q,\ \ p_x = p_r,\ \ p_y = r \tag{84}$$

in the equations of motion of Model I (Eq. (30)). It is trivial to show that Eq. (30) becomes

$$\begin{aligned}
\dot{q}^{(n)} &= H_{nn} p_q^{(n)} + \sum_{m\neq n} H_{nm} p_q^{(m)} \\
\dot{p}_q^{(n)} &= -H_{nn} q^{(n)} - \sum_{m\neq n} H_{nm} q^{(m)} \\
\dot{p}_r^{(n)} &= -H_{nn} r^{(n)} - \sum_{m\neq n} H_{nm} r^{(m)} \\
\dot{r}^{(n)} &= H_{nn} p_r^{(n)} + \sum_{m\neq n} H_{nm} p_r^{(m)}
\end{aligned} \quad . \tag{85}$$

Note that the equations of motion for $\left\{ r^{(n)}, p_r^{(n)} \right\}$ are identical to those of $\left\{ q^{(n)}, p_q^{(n)} \right\}$. If the initial condition is chosen as

$$r^{(n)}(0) = q^{(n)}(0),\quad p_r^{(n)}(0) = p_q^{(n)}(0) \ \ , \tag{86}$$

then two of the four variables in Eq. (85) are redundant. Make the transformation

$$\begin{aligned}
x^{(n)} &= \frac{q^{(n)} + r^{(n)}}{\sqrt{2}} \\
p^{(n)} &= \frac{p_q^{(n)} + p_r^{(n)}}{\sqrt{2}}
\end{aligned} \quad . \tag{87}$$

Eq. (85) then reduces to Eq. (83), the equations of motion for the TDSE in the Cartesian phase space. That is, Model I is also in principle an exact mapping model of the TDSE onto the Cartesian phase space, although a different analogy is employed for developing it.

Similarly, it is trivial to verify that the equations of motion of any model in Models III-VI

are the same as Eq. (83) of the TDSE, when the initial condition

$$x^{(n)}(0) = p_y^{(n)}(0), \quad y^{(n)}(0) = -p_x^{(n)}(0) \tag{88}$$

is employed.

Although these different classical mapping models for the multi-state Hamiltonian operator [Eq. (5)] are generated from different analogies as shown in Section III 1-4, all the six distinct models (Models I-VI) are equivalent expressions of the TDSE when their initial conditions are carefully constructed.

Finally, we note that *not* all mapping models for equivalent expressions of the quantum multi-state Hamiltonian operator [Eq. (5)] are exact. For instance, the spin mapping model of Cotton and Miller[3] and the semiclassical mapping model of Swenson *et al.*[14] do not lead to exact results for Eq. (5). The unified framework proposed in the paper presents a systematic approach to construct classical mapping models in the Cartesian space for the multi-state Hamiltonian operator which are able to produce exact results.

## Appendix C: Algorithms and numerical examples

Below we show the algorithms for Models I-VI and then test them with a 3-state model.

### 1. Algorithms for Models I-VI

1) Model I

A symplectic algorithm for propagating the trajectory through a time interval $\Delta t$ for Eq. (30) is

$$
\begin{aligned}
x^{(n)} &\leftarrow x^{(n)} - \frac{\Delta t}{2}\left(H_{nn} y^{(n)} + \sum_{m \neq n} H_{nm} y^{(m)}\right) \\
p_x^{(n)} &\leftarrow p_x^{(n)} - \frac{\Delta t}{2}\left(H_{nn} p_y^{(n)} + \sum_{m \neq n} H_{nm} p_y^{(m)}\right) \\
y^{(n)} &\leftarrow y^{(n)} + \Delta t\left(H_{nn} x^{(n)} + \sum_{m \neq n} H_{nm} x^{(m)}\right) \\
p_y^{(n)} &\leftarrow p_y^{(n)} + \Delta t\left(H_{nn} p_x^{(n)} + \sum_{m \neq n} H_{nm} p_x^{(m)}\right) \\
x^{(n)} &\leftarrow x^{(n)} - \frac{\Delta t}{2}\left(H_{nn} y^{(n)} + \sum_{m \neq n} H_{nm} y^{(m)}\right) \\
p_x^{(n)} &\leftarrow p_x^{(n)} - \frac{\Delta t}{2}\left(H_{nn} p_y^{(n)} + \sum_{m \neq n} H_{nm} p_y^{(m)}\right)
\end{aligned} \quad . \qquad (89)
$$

If the initial state is $|n\rangle$, i.e., the occupation number of state $|n\rangle$ is 1 while those of the other states are 0,

$$
\begin{aligned}
& x^{(n)} p_y^{(n)} - y^{(n)} p_x^{(n)} = 1 \\
& x^{(m)} p_y^{(m)} - y^{(m)} p_x^{(m)} = 0 \quad (m \neq n)
\end{aligned} \quad , \qquad (90)
$$

the initial condition at time $t = 0$ is then constructed as

$$
\begin{aligned}
& x^{(n)}(0) = \cos\theta \\
& y^{(n)}(0) = \sin\theta \\
& p_x^{(n)}(0) = -\sin\theta \\
& p_y^{(n)}(0) = \cos\theta \\
& x^{(m)}(0) = 0 \\
& y^{(m)}(0) = 0 \\
& p_x^{(m)}(0) = 0 \\
& p_y^{(m)}(0) = 0 \qquad (m \neq n)
\end{aligned} \quad . \qquad (91)
$$

Here $\theta$ can be any real number between 0 and $2\pi$ .

2) Model II

A symplectic algorithm for propagating the trajectory through a time interval $\Delta t$ for Eq. (41) is

$$\begin{aligned} p^{(n)} &\leftarrow p^{(n)} - \frac{\Delta t}{2}\left(H_{nn}x^{(n)} + \sum_{m\neq n} H_{nm}x^{(m)}\right) \\ x^{(n)} &\leftarrow x^{(n)} + \Delta t\left(H_{nn}p^{(n)} + \sum_{m\neq n} H_{nm}p^{(m)}\right) \\ p^{(n)} &\leftarrow p^{(n)} - \frac{\Delta t}{2}\left(H_{nn}x^{(n)} + \sum_{m\neq n} H_{nm}x^{(m)}\right) \end{aligned} \quad . \tag{92}$$

If the initial state is $|n\rangle$, the occupation number representation is

$$\begin{aligned} &\frac{1}{2}\left(\left(x^{(n)}\right)^2 + \left(p^{(n)}\right)^2\right) = 1 \\ &\frac{1}{2}\left(\left(x^{(m)}\right)^2 + \left(p^{(m)}\right)^2\right) = 0 \quad (m \neq n) \end{aligned} \quad . \tag{93}$$

The initial condition at time $t = 0$ is then constructed as

$$\begin{aligned} &x^{(n)}(0) = \sqrt{2}\cos\theta \\ &p^{(n)}(0) = \sqrt{2}\sin\theta \\ &x^{(m)}(0) = 0 \\ &p^{(m)}(0) = 0 \qquad (m \neq n) \end{aligned} \quad , \tag{94}$$

Here $\theta$ can be any real number between 0 and $2\pi$ .

3) Model III

A symplectic algorithm for propagating the trajectory through a time interval $\Delta t$ for Eq. (50) is then proposed as

$$\begin{aligned} &\text{Step 1: propagate } \left\{x^{(n)}, p_x^{(n)}\right\} \text{ for a half time interval } \frac{\Delta t}{2} \\ &\text{Step 2: propagate } \left\{y^{(n)}, p_y^{(n)}\right\} \text{ for a time interval } \Delta\text{t} \\ &\text{Step 3: propagate } \left\{x^{(n)}, p_x^{(n)}\right\} \text{ for another half time interval } \frac{\Delta t}{2} \end{aligned} \quad . \tag{95}$$

Step 1 and Step 3 share the same procedure

$$
\begin{aligned}
&p_x^{(n)} \leftarrow p_x^{(n)} - \frac{\Delta t}{8} H_{nn} x^{(n)} \\
&\begin{cases} x^{(n)} \leftarrow x^{(n)} - \dfrac{\Delta t}{4}\left( \dfrac{1}{2} H_{nn} y^{(n)} + \sum_{m \neq n} H_{nm} y^{(m)} \right) \\ p_x^{(n)} \leftarrow p_x^{(n)} - \dfrac{\Delta t}{4}\left( \dfrac{1}{2} H_{nn} p_y^{(n)} + \sum_{m \neq n} H_{nm} p_y^{(m)} \right) \end{cases} \\
&x^{(n)} \leftarrow x^{(n)} + \frac{\Delta t}{4} H_{nn} p_x^{(n)} \\
&\begin{cases} x^{(n)} \leftarrow x^{(n)} - \dfrac{\Delta t}{4}\left( \dfrac{1}{2} H_{nn} y^{(n)} + \sum_{m \neq n} H_{nm} y^{(m)} \right) \\ p_x^{(n)} \leftarrow p_x^{(n)} - \dfrac{\Delta t}{4}\left( \dfrac{1}{2} H_{nn} p_y^{(n)} + \sum_{m \neq n} H_{nm} p_y^{(m)} \right) \end{cases} \\
&p_x^{(n)} \leftarrow p_x^{(n)} - \frac{\Delta t}{8} H_{nn} x^{(n)}
\end{aligned}
\quad , \qquad (96)
$$

and Step 2 reads

$$
\begin{aligned}
&p_y^{(n)} \leftarrow p_y^{(n)} - \frac{\Delta t}{4} H_{nn} y^{(n)} \\
&\begin{cases} y^{(n)} \leftarrow y^{(n)} + \dfrac{\Delta t}{2}\left( \dfrac{1}{2} H_{nn} x^{(n)} + \sum_{m \neq n} H_{nm} x^{(m)} \right) \\ p_y^{(n)} \leftarrow p_y^{(n)} + \dfrac{\Delta t}{2}\left( \dfrac{1}{2} H_{nn} p_x^{(n)} + \sum_{m \neq n} H_{nm} p_x^{(m)} \right) \end{cases} \\
&y^{(n)} \leftarrow y^{(n)} + \frac{\Delta t}{2} H_{nn} p_y^{(n)} \\
&\begin{cases} y^{(n)} \leftarrow y^{(n)} + \dfrac{\Delta t}{2}\left( \dfrac{1}{2} H_{nn} x^{(n)} + \sum_{m \neq n} H_{nm} x^{(m)} \right) \\ p_y^{(n)} \leftarrow p_y^{(n)} + \dfrac{\Delta t}{2}\left( \dfrac{1}{2} H_{nn} p_x^{(n)} + \sum_{m \neq n} H_{nm} p_x^{(m)} \right) \end{cases} \\
&p_y^{(n)} \leftarrow p_y^{(n)} - \frac{\Delta t}{4} H_{nn} y^{(n)}
\end{aligned}
\quad . \qquad (97)
$$

When the initial state is $|n\rangle$, i.e.,

$$
\begin{aligned}
&\frac{1}{4}\left( \left( x^{(n)} + p_y^{(n)} \right)^2 + \left( y^{(n)} - p_x^{(n)} \right)^2 \right) = 1 \\
&\frac{1}{4}\left( \left( x^{(m)} + p_y^{(m)} \right)^2 + \left( y^{(m)} - p_x^{(m)} \right)^2 \right) = 0 \quad (m \neq n)
\end{aligned}
\quad , \qquad (98)
$$

the initial condition at time $t = 0$ can be constructed the same as Eq. (91).

4) Model IV

A symplectic algorithm for propagating the trajectory through a time interval $\Delta t$ for Eq. (55)is

$$\begin{array}{ll} \text{Step 1: propagate } \left\{p_x^{(n)}, p_y^{(n)}\right\} \text{ for a half time interval } \frac{\Delta t}{2} & \\ \text{Step 2: propagate } \left\{x^{(n)}, y^{(n)}\right\} \text{ for a time interval } \Delta \text{t} & , \\ \text{Step 3: propagate } \left\{p_x^{(n)}, p_y^{(n)}\right\} \text{ for another half time interval } \frac{\Delta t}{2} & \end{array} \tag{99}$$

where either Step 1 or Step 3 reads

$$\begin{aligned} & p_x^{(n)} \leftarrow p_x^{(n)} - \frac{\Delta t}{8} H_{nn} p_y^{(n)} \\ & \begin{cases} p_x^{(n)} \leftarrow p_x^{(n)} - \frac{\Delta t}{4}\left(\frac{1}{2} H_{nn} x^{(n)} + \sum_{m \neq n} H_{nm} x^{(m)}\right) \\ p_y^{(n)} \leftarrow p_y^{(n)} - \frac{\Delta t}{4}\left(\frac{1}{2} H_{nn} y^{(n)} + \sum_{m \neq n} H_{nm} y^{(m)}\right) \end{cases} \\ & p_y^{(n)} \leftarrow p_y^{(n)} + \frac{\Delta t}{4} H_{nn} p_x^{(n)} \\ & \begin{cases} p_x^{(n)} \leftarrow p_x^{(n)} - \frac{\Delta t}{4}\left(\frac{1}{2} H_{nn} x^{(n)} + \sum_{m \neq n} H_{nm} x^{(m)}\right) \\ p_y^{(n)} \leftarrow p_y^{(n)} - \frac{\Delta t}{4}\left(\frac{1}{2} H_{nn} y^{(n)} + \sum_{m \neq n} H_{nm} y^{(m)}\right) \end{cases} \\ & p_x^{(n)} \leftarrow p_x^{(n)} - \frac{\Delta t}{8} H_{nn} p_y^{(n)} \end{aligned} \tag{100}$$

and Step 2 is

$$
\begin{aligned}
&x^{(n)} \leftarrow x^{(n)} - \frac{\Delta t}{4} H_{nn} y^{(n)} \\
&\begin{cases}
x^{(n)} \leftarrow x^{(n)} + \dfrac{\Delta t}{2}\left( \dfrac{1}{2} H_{nn} p_x^{(n)} + \sum_{m \neq n} H_{nm} p_x^{(m)} \right) \\
y^{(n)} \leftarrow y^{(n)} + \dfrac{\Delta t}{2}\left( \dfrac{1}{2} H_{nn} p_y^{(n)} + \sum_{m \neq n} H_{nm} p_y^{(m)} \right)
\end{cases} \\
&y^{(n)} \leftarrow y^{(n)} + \frac{\Delta t}{2} H_{nn} x^{(n)} \\
&\begin{cases}
x^{(n)} \leftarrow x^{(n)} + \dfrac{\Delta t}{2}\left( \dfrac{1}{2} H_{nn} p_x^{(n)} + \sum_{m \neq n} H_{nm} p_x^{(m)} \right) \\
y^{(n)} \leftarrow y^{(n)} + \dfrac{\Delta t}{2}\left( \dfrac{1}{2} H_{nn} p_y^{(n)} + \sum_{m \neq n} H_{nm} p_y^{(m)} \right)
\end{cases} \\
&x^{(n)} \leftarrow x^{(n)} - \frac{\Delta t}{4} H_{nn} y^{(n)}
\end{aligned}
\quad . \qquad (101)
$$

When the initial state is $|n\rangle$, the occupation number representation is the same as Eq. (98) and the corresponding initial condition at time $t = 0$ is set the same as Eq. (91).

Similar algorithms can be proposed for Models V and VI, as done for Models III and IV. We do not repeat the procedure. (We also note that Models III-VI can also employ the same algorithm as Eq. (85) when Eq. (91) is the initial condition at time $t = 0$ although the performance is not as good for these models.) It should be emphasized that a single trajectory is sufficient for obtaining dynamics of underlying degrees of freedom in any classical mapping model of Models I-VI. (Note that $\{H_{nm}\}$ in Eq. (1) are time-independent in the present paper.)

The simplectic algorithms of Models I-VI can also be viewed as robust numerical integrators for solving the TDSE.

**2. Three-state model**

A simple but non-trivial case is a 3-state system. The Hamiltonian operator is given by

$$\hat{H} = \sum_{n=1}^{3} H_{nn} |n\rangle\langle n| + \sum_{m<n} H_{mn} \left( |m\rangle\langle n| + |n\rangle\langle m| \right) \quad . \tag{102}$$

The diagonal terms in Eq. (102) are

$$H_{11} = 10, \quad H_{22} = 7, \quad H_{33} = 2 \quad , \tag{103}$$

and the off-diagonal ones are

$$H_{12} = H_{13} = H_{23} = \lambda \quad , \tag{104}$$

where $\lambda$ is a parameter. Its value is set to be 0.02, 0.2, 2, or 20 to cover from the weak coupling regime to the strong coupling domain.

The initial state is chosen to be $|1\rangle$ (or the initial density is $|1\rangle\langle 1|$). The corresponding initial condition in Model I or III-VI is then given by Eq. (91), while that in Model II is constructed as Eq. (94). The results of any one of the six classical mapping models are independent of the value of $\theta$ in Eq. (91) or Eq. (94). Figs. 1-4 present the population of each basis state of $\{ |n\rangle \}$ as a function of time produced by the six models. Comparison to the quantum results shows that each of the six models leads to exact population dynamics in all test cases from the weak coupling regime to the strong coupling domain.

The time interval of the trajectory propagation in Model I or III-VI is often larger than that in Model II for achieving the same accuracy, especially in the weak coupling regime. For example, Fig. 5 shows that the time interval in Model I is about 100 times of that in Model II for achieving the same accuracy when the coupling is $\lambda = 0.02$ in Eq. (104). While the six different classical mapping models perform similarly in the strong coupling domain, Model I and Models III-VI demonstrate better numerical performance than Model II in the weak coupling regime.

Finally, we compare the spin mapping model (SPM) to Model VI. While both mapping models are developed from the equivalent expressions [Eq. (58), (61) or Eq. (63)] of the multi-state Hamiltonian operator in quantum mechanics, different strategies are employed. SPM replaces the spin components by its classical angular momentum counterparts [i.e., Eq. (65)]. The equations of motion for SPM are given by Eq. (68). Consider a pure state as the initial condition, e.g. the *i*-th state is occupied. The classical mapping model [Eq. (68)] employs $m^{(i)}=1/2$ for the occupied state and $m^{(j)}=-1/2$ for unoccupied states $(j\neq i)$ in Eq. (65), while the angle variable $q^{(i)}$ or $q^{(j)}$ can take any real value between 0 and $2\pi$. A single trajectory is *not* sufficient for obtaining meaningful population dynamics results in SPM. Instead an ensemble of trajectories with different initial conditions for $q^{(i)}$ or $q^{(j)}$ are employed in SPM. (This is different from Models I-VI where only one trajectory is sufficient in the classical limit.) The initial values of the angle variables can be either uniformly or randomly chosen between 0 and $2\pi$, as long as enough number of trajectories are used for obtaining converged results. Fig. 6 shows that SPM only works reasonably well in the weak coupling regime, but becomes progressively worse as the coupling terms get stronger. For comparison, Model VI reproduces exact results in any coupling regimes.

One can further employ Cotton and Miller's quasiclassical approach for SPM[3] for the 3-state Hamiltonian. While the initial values of the angle variables are (either uniformly or randomly) chosen between 0 and $2\pi$, those of the action variables are similarly chosen within a distance $(\sqrt{3}-1)/2\approx 0.366$ of $1/2$ or $-1/2$ for occupied or unoccupied states, respectively. Consider the case of Fig. 6c as an example, where the coupling between any two states is $\lambda = 2$ in Eq. (104). Fig. 7 demonstrates that the quasiclassical approach for SPM does not work well either in the strong coupling regime. This then explains well why the quasiclassical approach of SPM is less accurate than that of the Meyer-Miller mapping model as demonstrated in Ref. [3].

**Figure Captions**

**Fig. 1** (Color). Population dynamics of the 3-state Hamiltonian system. (Its parameters given by Eq. (103) and Eq. (104).) The initial state is $|1\rangle$. The coupling between any two states is $\lambda = 0.02$ in Eq. (104). (a) Population of state $|1\rangle$ (as a function of time). (b) Population of state $|2\rangle$. (c) Population of state $|3\rangle$. Solid line: exact results. Solid squares: results of Model I. Solid triangles: results of Model II. Solid rhombuses: results of Model III. Solid circles: results of Model IV. Crosses: results of Model V. Hollow squares: results of Model VI. ($\hbar = 1$)

**Fig. 2** (Color). As in Fig. 1. The coupling between any two states is $\lambda = 0.2$.

**Fig. 3** (Color). As in Fig. 1. The coupling between any two states is $\lambda = 2$.

**Fig. 4** (Color). As in Fig. 1. The coupling between any two states is $\lambda = 20$.

**Fig. 5** (Color). Comparison between Model I and Model II for the population of state $|1\rangle$ as a function of time for the 3-state Hamiltonian system in the weak coupling regime. (Its parameters given by Eq. (103) and Eq. (104).) The initial state is $|1\rangle$. The coupling between any two states is $\lambda = 0.02$ in Eq. (104). Solid line (black): exact results. Dotted line (blue): Model I (time interval $\Delta t = 10^{-3}$). Dashed line (red): Model II (time interval $\Delta t = 10^{-3}$). Dot-dashed line (green): Model II (time interval $\Delta t = 10^{-4}$). Short-dashed line (purple): Model II (time interval $\Delta t = 10^{-5}$). Panel (b) is a blow-up of Panel (a) for the time regime $[2.9, 3.4]$.

**Fig. 6** (Color). Comparison between Model VI and the spin mapping model for the population of state $|1\rangle$ as a function of time for the 3-state Hamiltonian system. (Its parameters given by Eq. (103) and Eq. (104).) The initial state is $|1\rangle$. (a) The coupling

between any two states is $\lambda = 0.02$ in Eq. (104). (b) $\lambda = 0.2$. (c) $\lambda = 2$. (d) $\lambda = 20$. Solid line (black): exact results. Dotted line (blue): Model VI. Dashed line (red): Spin mapping model (SPM).

**Fig. 7** (Color). Comparison between classical and quasiclassical dynamics for SPM for the population of state $|1\rangle$ as a function of time for the 3-state Hamiltonian system. (Its parameters given by Eq. (103) and Eq. (104).) The initial state is $|1\rangle$. The coupling between any two states is $\lambda = 2$ in Eq. (104). Solid line (black): exact results. Dotted line (red): classical dynamics for SPM. Dashed line (blue): quasiclassical dynamics for SPM[3].

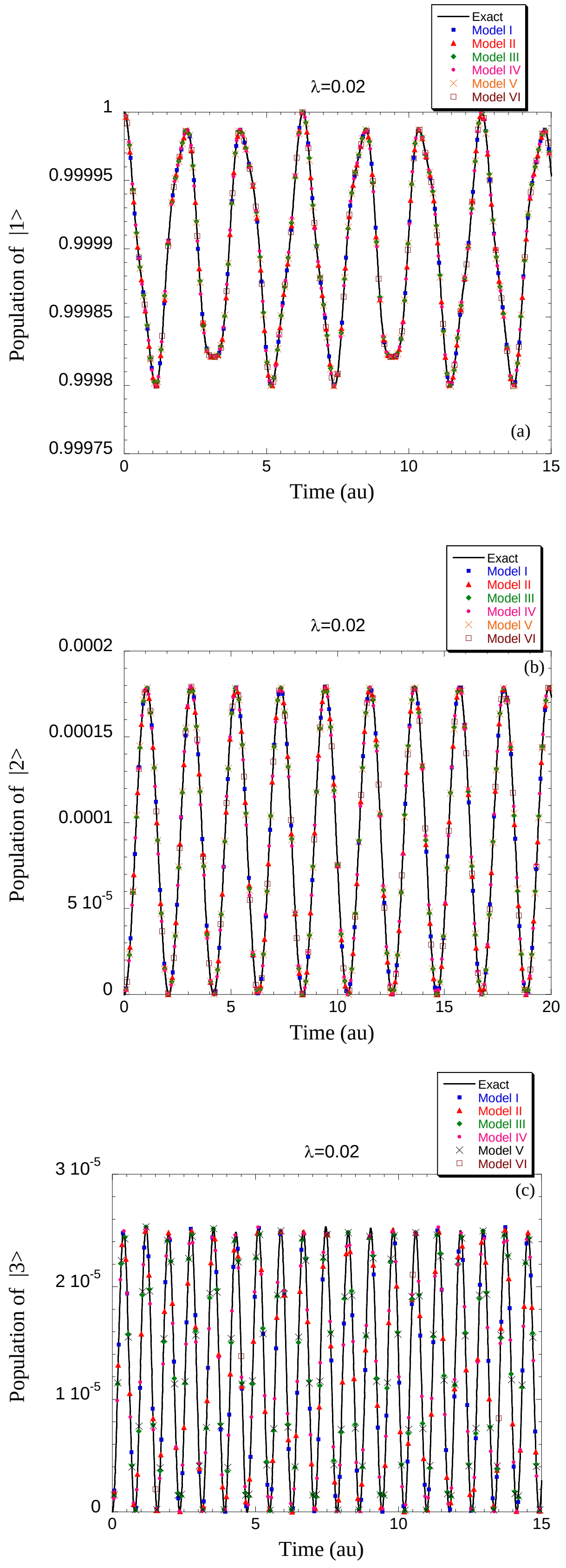

**Fig. 1**

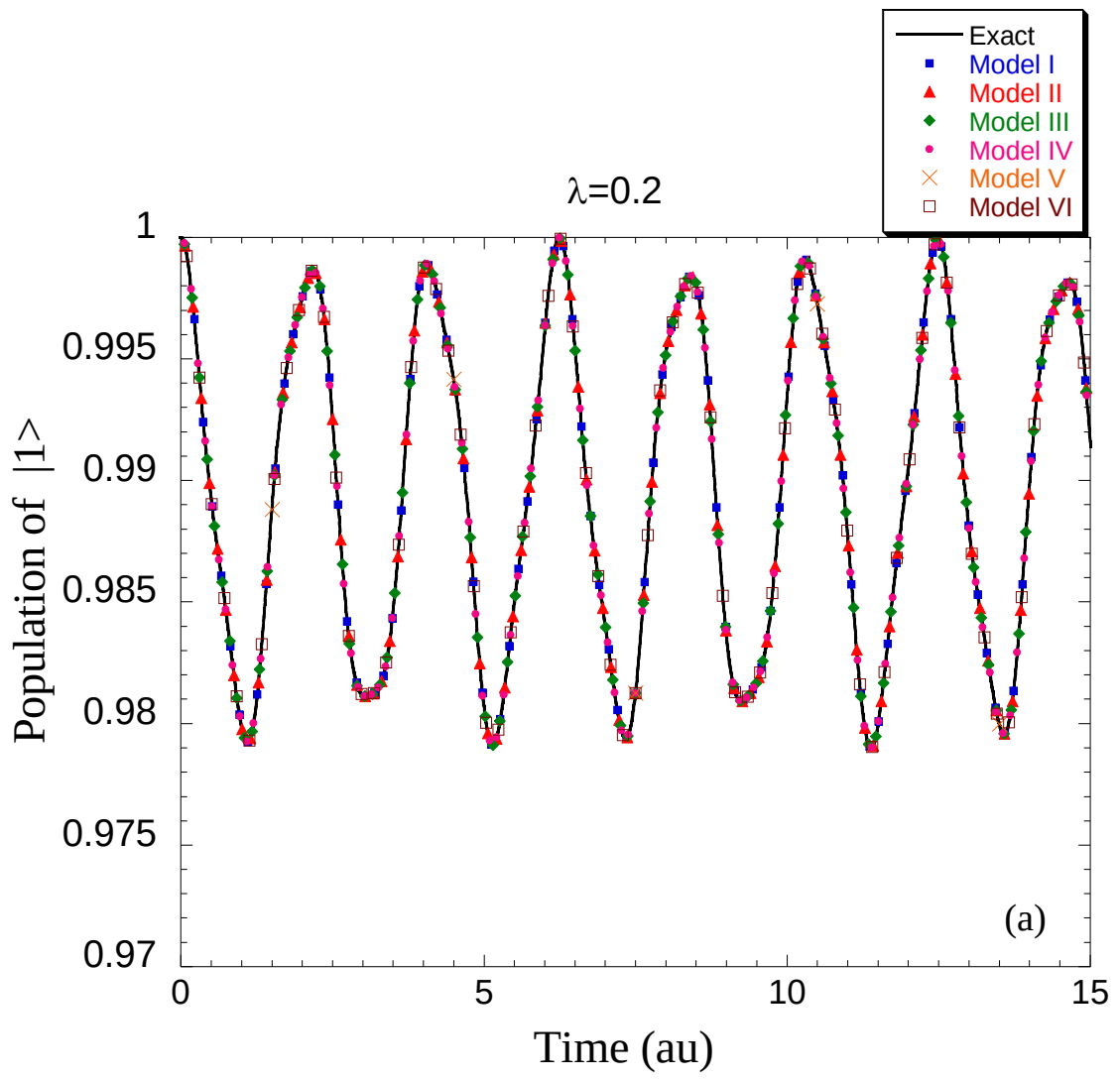

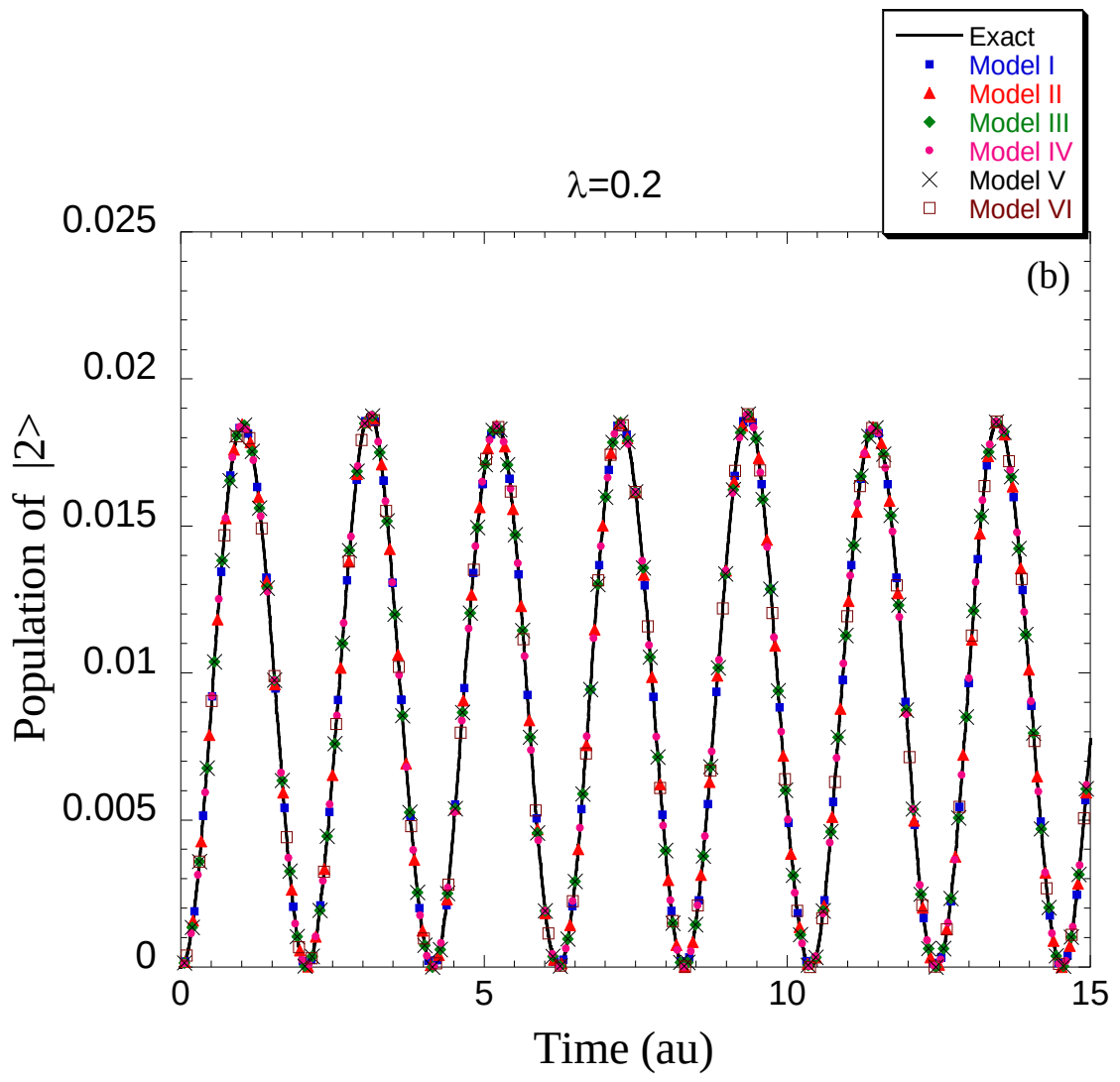

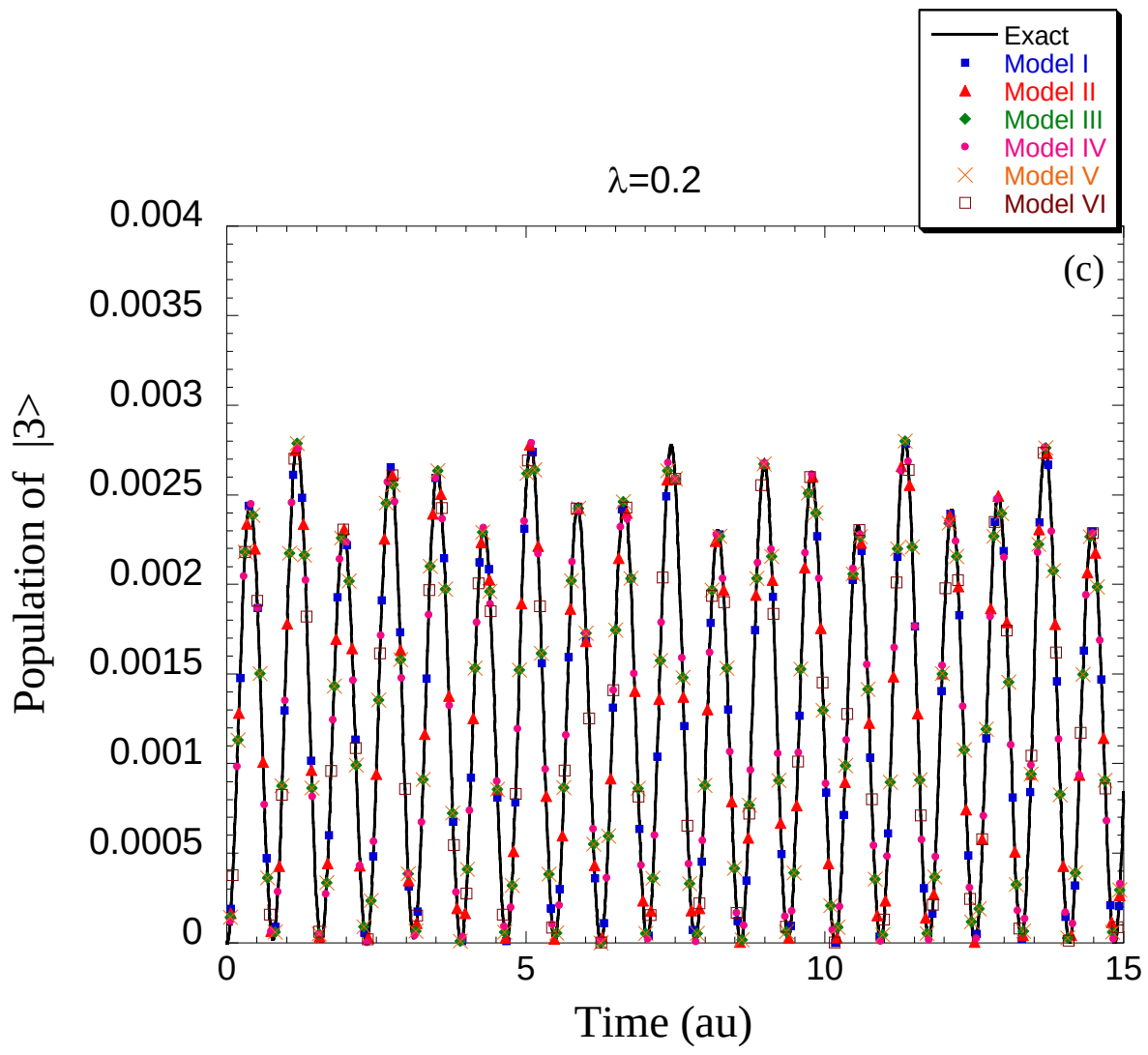

**Fig. 2**

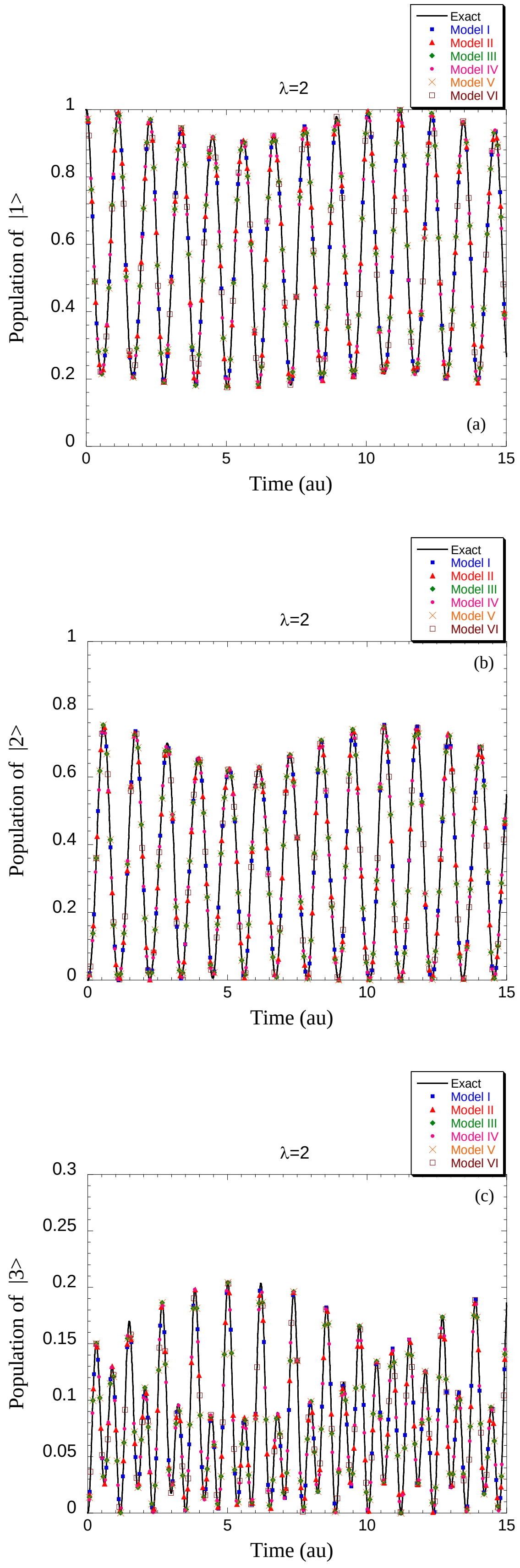

**Fig. 3**

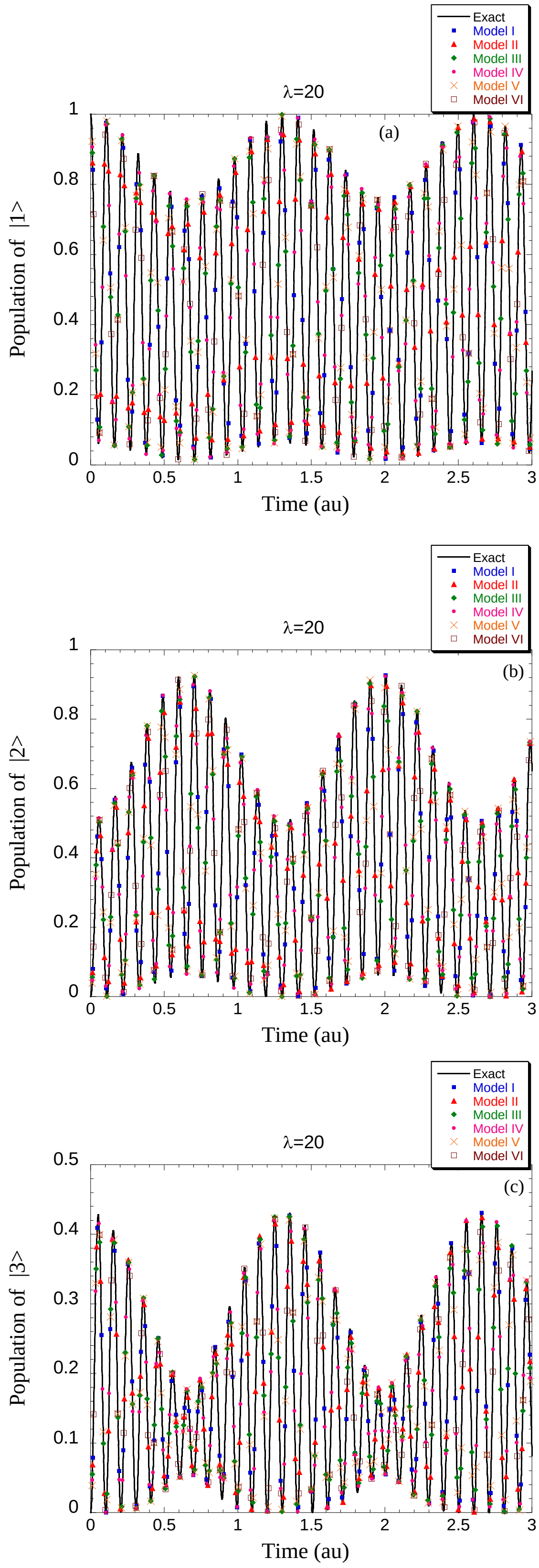

**Fig. 4**

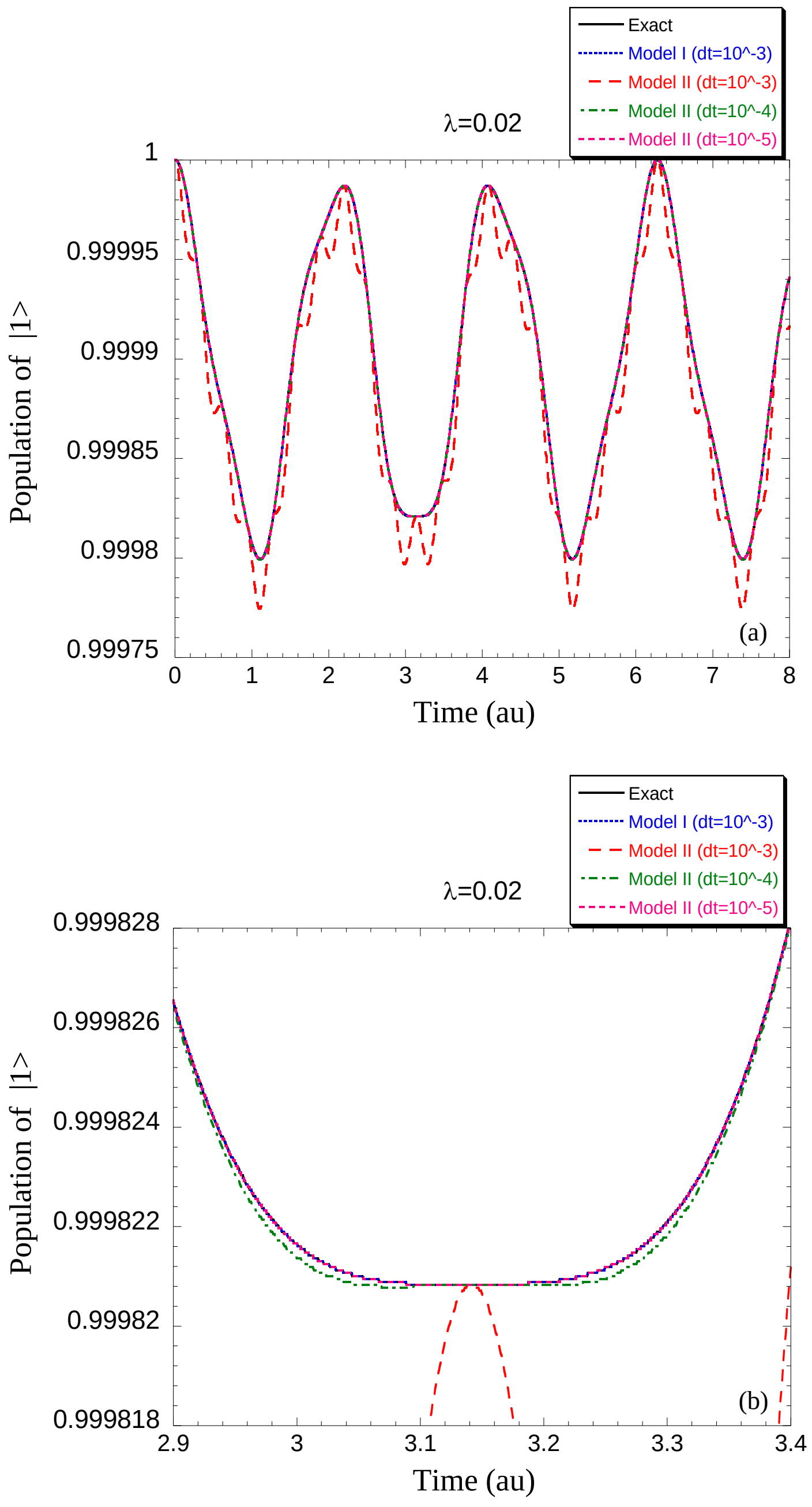

**Fig. 5**

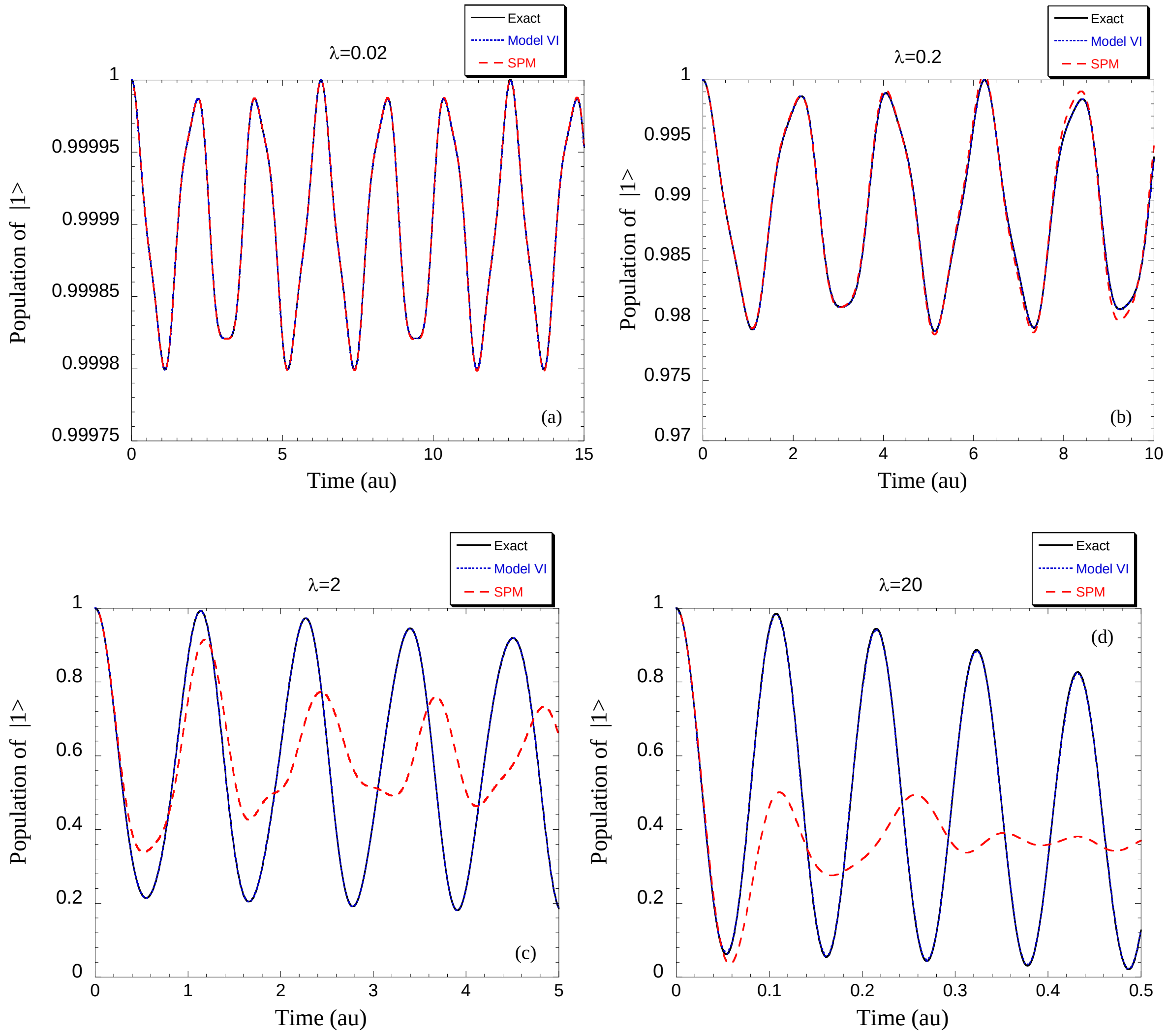

**Fig. 6**

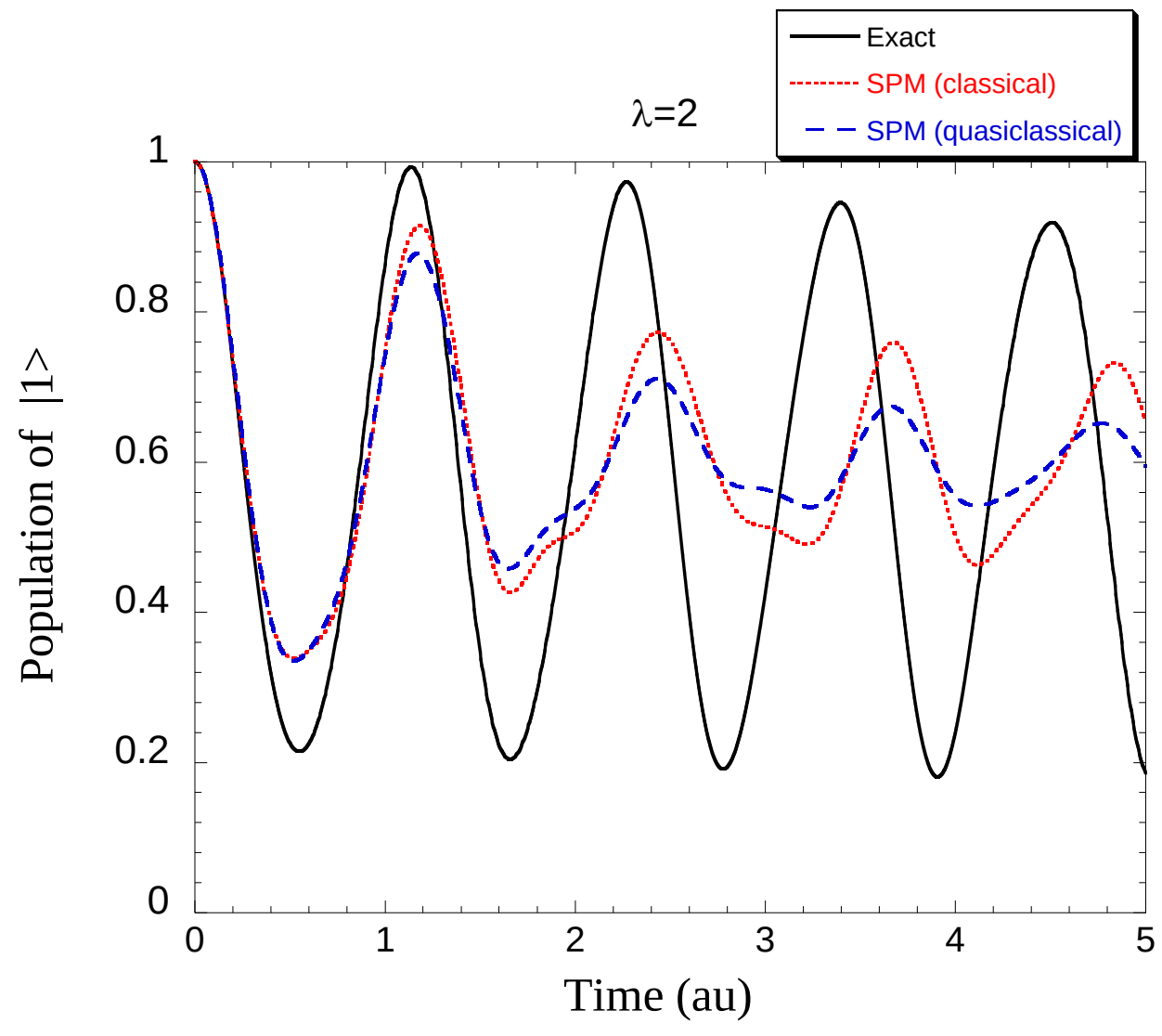

**Fig. 7**